\documentclass[a4paper,12pt]{extarticle}
\usepackage[utf8]{inputenc}
\usepackage[singlespacing]{setspace}
\pdfoutput=1
\usepackage{jcappub} % for details on the use of the package, please
\usepackage[top=20mm,left=20mm,right=20mm,bottom=25mm,foot=10mm]{geometry}
\usepackage{subfig}
\usepackage[dvipsnames]{xcolor}
                     
\newcommand{\eps}{\epsilon}
\newcommand{\Pmax}{p_\text{max}} 	
\newcommand{\Hint}{H_{int}}
\newcommand{\kbar}{\bar{k}}

\newcommand{\shapecor}{\mathcal{S}}
\newcommand{\ampcor}{\mathcal{A}}
\newcommand{\totalcor}{\mathcal{E}}

\newcommand{\Lbasic}{\mathcal{P}_0}
\newcommand{\Linvk}{\mathcal{P}_1}
\newcommand{\Lnsinv}{\mathcal{P}^{n_s}_1}
\newcommand{\Lnsboth}{\mathcal{P}^{n_s}_{01}}

\newcommand{\Fbasic}{\mathcal{F}_0}
\newcommand{\Finvk}{\mathcal{F}_1}

\newcommand{\quadpot}{V_{\phi^2}(\phi)}

\newcommand{\threeqs}{q_p(k_1)\,q_r(k_2)\,q_s(k_3)}

\newcommand{\kmin}{{k_\text{min}}}
\newcommand{\kmax}{{k_\text{max}}}
\newcommand{\fnl}{f_{NL}}
\newcommand{\fnllocal}{f^{local}_{NL}}
\newcommand{\fnlequil}{f^{equil}_{NL}}
\newcommand{\fnlortho}{f^{ortho}_{NL}}

\usepackage[T1]{fontenc} % if needed

\title{Probing Inflation with Precision Bispectra}

\author{
    Philip Clarke and E. P. S. Shellard
}
\affiliation{Centre for Theoretical Cosmology, DAMTP, University of Cambridge, Cambridge CB3 0WA, United Kingdom}

\emailAdd{pc559@cam.ac.uk}

\abstract{
    Calculating the primordial bispectrum predicted by a model of inflation and
    comparing it to what we see in the sky is very computationally intensive, necessitating layers
    of approximations and limiting the models which can be constrained.  Exploiting the inherent separability
    of the tree level in-in formalism using expansions in separable basis functions provides a means by which to
    obviate some of these difficulties. Here, we develop this approach further into a practical and efficient
    numerical methodology which can be applied to a much wider and more complicated range of bispectrum
    phenomenology, making an important step forward towards observational pipelines which can directly
    confront specific models of inflation.  We describe a simple augmented Legendre polynomial basis
    and its advantages, then test the method on single-field inflation models with non-trivial phenomenology,
    showing that our calculation of these coefficients is fast and accurate to high orders.
}

\begin{document}
\maketitle
\flushbottom
\newpage

\section{Introduction}\label{sec:intro}
%\subsection{The bispectrum}
The primordial bispectrum is one of the main
characteristics used to distinguish between models of inflation. While it is well
known that the physics of inflation must have been extremely close
to linear, and the initial seeds of structure it laid down
very close to Gaussian, there is expected to have been some level of coupling
between the Fourier modes of the perturbations.
In the simplest example of an inflation model this is
expected to be unobservable~\cite{Maldacena},
but the possibility remains that inflation was driven by
more complex physics that may have left an observable imprint on our universe today.
Some models of inflation have interactions that predict non-Gaussian
correlations at observable levels. Ways this can happen include
self-interactions~\cite{px_burrage,dbi_in_the_sky},
interactions between multiple fields~\cite{Byrnes_2010},
sharp features~\cite{adshead}
and periodic features~\cite{flauger_pajer_resonant}.
However, constraining such imprints is extremely difficult observationally.
Even once the data has been obtained, using existing methods it is
extremely computationally intensive to translate this into constraints
on specific inflation scenarios. Much progress has been made by course-graining
the model space into a small number of approximate templates,
and leveraging the simplifying characteristic of separability
with respect to the three parameters of the bispectrum~\cite{Komatsu_2005, Munchmeyer_2014}.

The primordial bispectrum is the Fourier equivalent of the
three-point correlator of the primordial curvature perturbation.
If this field is Gaussian, the bispectrum vanishes, so
it is a valuable measure of the interactions in play during inflation.
If some inflation model predicts a bispectrum that is sufficiently well approximated by
the standard separable templates, the constraints on those standard templates
can be translated into constraints on the parameters of the model.
The fact that all primordial templates estimated thus far from the CMB
are consistent with zero has already provided such constraints
in certain scenarios~\cite{Planck_NG_2015, Planck_NG_2018}.
With this high-precision \textit{Planck} data, and data from forthcoming experiments
such as the Simons Observatory (SO)~\cite{simons}
and CMB-S4~\cite{abazajian2016cmbs4},
robust pipelines must be developed to circumvent the computational difficulties and
extract the maximum amount of information possible.
Due to the nature of bispectrum estimation in the CMB and
LSS~\cite{lss_baldauf,lss_karagiannis,chen_future_lss,Scoccimarro_2012}
constraining an arbitrary template is difficult.
Our aim in this work is to develop the inflationary part
of a pipeline to allow to efficiently test a much broader range of models.
In this work, we explore shapes arising from tree-level effects in single field models.
We do this numerically, allowing quantitative results for a broad
range of models, and avoiding extra approximations.
Our general aim is to apply the modal philosophy of~\cite{FergShell_1,FergShell_2,FergShell_3}
to calculating primordial bispectra.
This modal philosophy is a flexible method that has broadened the range of constrained
bispectrum templates, by expanding them in a carefully chosen basis.
The Modal estimator is thus capable of constraining
non-separable templates, while the KSW estimator cannot.
In this work we exploit the intrinsic separability of the
tree-level in-in formalism to apply these methods at the level of inflation.
Expressing the primordial bispectrum in a separable
basis expansion leads to vast increases in efficiency both at the primordial
and late-universe parts of the calculation.
The main advantage is that expressing the primordial shape function
in this way reduces the process of bispectrum estimation in the CMB to a
cost which is large, but need only be paid once per basis,
not per scenario.
A proof of concept of this approach at the primordial level was presented in~\cite{Funakoshi},
and the details of the bispectrum estimation part will be detailed in~\cite{Sohn_2020}.
We go beyond the work of~\cite{Funakoshi} both in developing the choice of basis
(the feasibility of the method depending vitally on the chosen basis
achieving sufficiently fast convergence in a broad range of interesting models)
and in the methods we use to allow us to go to much higher order in our modal expansion,
allowing us to apply the method to feature bispectra for the first time.

The paper is organised as follows. In section~\ref{sec:review} we present brief reviews
of the various parts of the pipeline that connects inflation scenarios to observations
through the bispectrum.
We review the usual paradigm of bispectrum estimation in the CMB,
and the motivation for separable bispectra. We review the in-in formalism,
for calculating the tree level bispectrum for a given model of inflation.
We review $P(X,\phi)$ models of inflation as an example, and
some of the usual approximate bispectrum templates
that we aim to bypass.
We will draw our validation scenarios from these models.
We discuss previous numerical codes for
calculating the primordial bispectrum $k$-configuration by $k$-configuration,
which contrasts our separable basis expansion.
We review the previous work in achieving separability through modal expansions
in~\cite{Funakoshi},
and we discuss methods of testing
numerical bispectrum results, defining our relative difference measurement.
In section~\ref{sec:methods} we present our methods.
Since the paradigm we aim to present is only viable if we can find a basis
that can efficiently represent a wide variety of bispectra,
we begin with this vital discussion. We discuss the effects of the
non-physical $k$-configurations on the convergence of our expansion on
the tetrapyd, and present an efficient basis.
Then, we recast the usual in-in calculation into an explicitly separable form,
in terms of an expansion in an arbitrary basis,
and detail our methods for carefully calculating the coefficients to high order.
In section~\ref{sec:validation} we validate our methods and implementation
on inflation scenarios with varied features from the literature,
and we finish with a discussion of future work in section~\ref{sec:future}.

\section{Inflationary bispectra and observations}\label{sec:review}
\subsection{Bispectrum estimation}\label{sec:rev_bis_est}
The bispectrum, like the power spectrum, is a quantity that describes
the statistical distribution of which our universe is only one realisation.
We use this one sky we have access to to estimate the amplitude of
particular bispectrum templates,
and use these estimates to constrain inflationary physics; 
see~\cite{astro2020_features,astro2020_png} for recent reviews.
There are two parts to the pipeline of bispectrum estimation.
Firstly, calculating the primordial bispectrum at the end of some inflation scenario,
and then calculating the effect this bispectrum
has on some appropriate observable today.
One well-developed example is
the bispectrum of temperature fluctuations in the CMB, which uses transfer functions
to evolve and project the primordial bispectrum onto our sky.
In principle, this is the same process as power spectrum estimation.
However, for the bispectrum the computational challenge is far greater,
requiring both compute-intensive and large in-memory components.

As a result of this complexity, this second step is computationally impractical for generic primordial bispectra.
Progress can be made by finding an approximation to the primordial shape
that is separable, and using this simplification
to make the calculation tractable
through the KSW estimator~\cite{Komatsu_2005, Munchmeyer_2014}.
For example, one may find that a particular inflation scenario generates
a primordial bispectrum with a high correlation with some standard shape,
then look at how well that standard shape is constrained by the CMB.
The modal decomposition method of~\cite{FergShell_1,FergShell_2,FergShell_3}
leveraged these simplifications in a more structured way
for generic bispectra, broadening the range of constrained models.

The measure of non-Gaussianity in the CMB that is
most usually quoted is $\fnl$, referring to $\fnllocal$.
This number describes how well a particular template, the local template,
describes the correlations in the CMB;
this template is used as a proxy for the class of inflation models that produce similar bispectra.
Similar quantities for the equilateral and orthogonal templates are also
commonly quoted.
In addition to broadening the range of constrained models through increases in efficiency,
the modal decomposition method of~\cite{FergShell_1,FergShell_2,FergShell_3}
allows to go beyond this paradigm, efficiently constraining inflationary bispectra in the CMB using
all of the shape information; essentially constraining an $\fnl$
specific to a given bispectrum. This bypasses the approximation step at the level of the templates,
of finding a separable approximation to the primordial bispectrum.
In this work, our numerical methods remove the need for some of the approximations
made before this, during inflation, directly linking the parameters of the inflation scenario
with the relevant observable.
In addition to this improvement in accuracy, calculating the modal decomposition
directly from the model of inflation is far more efficient 
than numerically calculating the bispectrum configuration by configuration.

The primordial bispectrum is usually written as:
\begin{align}
{
\left< \zeta_{\bf{k_1}}\zeta_{\bf{k_2}}\zeta_{\bf{k_3}} \right>
= (2\pi)^3\delta^{(3)}(\mathbf{k_1}+\mathbf{k_2}+\mathbf{k_3})B(k_1,k_2,k_3)
}
\end{align}
where $\zeta_{\bf{k}}$ is a Fourier mode of the standard gauge invariant curvature perturbation.
The delta function comes from demanding statistical homogeneity;
demanding statistical isotropy restricts the remaining dependence to the magnitudes of the vectors.
We denote the magnitude of $\mathbf{k_i}$ as $k_i$.
This leaves us with a function of three parameters,
$k_1,k_2,k_3$.
It is useful to define the dimensionless shape function:
\begin{align}\label{shapefn}
{
    S(k_1,k_2,k_3) = (k_1k_2k_3)^2B(k_1,k_2,k_3).
}
\end{align}
The bispectrum is only defined where the triangle condition
\begin{align}\label{triangle_condition}
    \mathbf{k_1}+\mathbf{k_2}+\mathbf{k_3} &= 0,
\end{align}
is satisfied, which implies that the triangle inequality must hold
\begin{align}\label{triangle_inequality}
    k_1+k_2 &\geq k_3~\text{and cyclic perms}.
\end{align}
The space of configurations we are interested in is therefore
reduced from the full cube $[\kmin,\kmax]^3$
to a tetrapyd (illustrated in figure~\ref{slice_explained}),
the intersection of that cube with the
tetrahedron that satisfies~\eqref{triangle_inequality}.
This has important implications that we will explore in
section~\ref{sec:choice_of_basis}.

The amplitude of a bispectrum shape is usually
quoted in terms of some $\fnl^F$ parameter.
We can schematically define $\fnl^F$ for some template $F$ as follows:
\begin{equation}\label{def:fnl}
{
B^{F}(k_1,k_2,k_3) = \fnl^{F}\times F(k_1,k_2,k_3)
}
\end{equation}
where $F$ contains the dependence on the $k$-configuration.
This definition coincides with the definitions of
$\fnllocal$, $\fnlequil$ and $\fnlortho$
when $F$ is (respectively) the local (see \eqref{local_shape}),
equilateral (see \eqref{equil_shape}) and orthogonal templates,
as defined in~\cite{Planck_NG_2015}.

If the shape function~\eqref{shapefn} has the form:
\begin{equation}\label{sepXYZ}
S(k_1,k_2,k_3) = X(k_1)Y(k_2)Z(k_3),
\end{equation}
or can be expressed as a sum of such terms,
it is called separable.
The link between the separability of the primordial bispectrum
and the reduced CMB bispectrum can be seen from the following expression:
\begin{equation}
\label{eq:reduced_cmb}
b^{X_1X_2X_3}_{l_1l_2l_3} = \left(\frac{2}{\pi}\right)^3\int_{0}^{\infty}drr^2\int_{\mathcal{V}_k}d^3k (k_1k_2k_3)^2 B_{\Phi}(k_1,k_2,k_3)\prod_{i=1}^{3}\left[j_{l_i}(k_ir)\Delta^{X_i}_{l_i}(k_i)\right],
\end{equation}
where we also see that if the primordial bispectrum is separable
then the overall dimension
of the calculation can be reduced from seven to five, 
since the spherical Bessel functions $j_{l_i}$ and the
transfer functions $\Delta_{l_i}$ already appear in a separable way.
This property can also be used to
efficiently generate non-Gaussian initial conditions
for simulations~\cite{Scoccimarro_2012}.

The numbers $\fnl^F$ are useful summary parameters.
From the data-side, they represent the result of
a complex and intensive process
of estimating the amplitude of the template $F$,
given some data. From the theory-side, one
can use them to take an inflation scenario and compare it
to that data, if one can find a standard template
with a high correlation with the shape resulting
from that scenario.
However, despite its usefulness, this paradigm does
have drawbacks. It acts as an information bottleneck,
losing some constraining power when one approximates
the real shape function by some standard template.
In particular, if one is interested in a feature model,
it may be be difficult to see how constraints on existing
features can be applied.

\subsection{The tree level in-in calculation}\label{sec:rev_inin}
The standard starting point for calculating
higher-order correlators for models of inflation is the in-in formalism~\cite{Maldacena,weinberg_in_in}.
The in-in formalism takes the time evolution of the interaction picture mode
functions as an input for calculating the bispectrum.
At tree-level, the in-in formalism gives us the
following expression:
\begin{align}
{
    \left< \zeta_{\mathbf{k_1}}(\tau)\zeta_{\mathbf{k_2}}(\tau)\zeta_{\mathbf{k_3}}(\tau) \right>
=-i\int_{-\infty(1-i\varepsilon)}^{\tau}d\tau'a(\tau')
    \left<0\lvert \zeta_{\mathbf{k_1}}(\tau)\zeta_{\mathbf{k_2}}(\tau)\zeta_{\mathbf{k_3}}(\tau)\Hint(\tau') \rvert0\right>+c.c\label{in-in}
}
\end{align}
where all the operators on the right-hand side are in the interaction picture
and $\Hint$ is the interaction Hamiltonian, containing terms cubic in $\zeta$.
From this calculation we obtain the dimensionless shape function $S(k_1,k_2,k_2)$,
defined in~\eqref{shapefn},
which is then used as input into~\eqref{eq:reduced_cmb}.
As an example, if one takes $\Hint\propto\dot{\zeta}^3$, this set-up can produce the standard EFT shape
\begin{align}\label{example_eft2}
    S(k_1, k_2, k_3) = \frac{k_1k_2k_3}{(k_1+k_2+k_3)^3}.
\end{align}

The central point, as noticed in~\cite{Funakoshi}, is that the
integrand of~\eqref{in-in} is intrinsically separable
in its dependence on $k_1$, $k_2$ and $k_3$, and that the time integral
can be done in such a way as to preserve this separability.
This intrinsic separability has clearly been lost in
the example in~\eqref{example_eft2},
but can be regained (to arbitrary precision) by approximating it
with a sum of separable terms. Our general aim will be to directly calculate
this sum for a broad range of inflation models.

We now briefly outline the set-up of the standard calculation.
The Lagrangian is expanded in the perturbations and used to obtain the Hamiltonian.
The Hamiltonian is split into $H_0$ and $\Hint$.
The first part is used to evolve the interaction picture fields, $\zeta_I$,
which we will simply refer to as $\zeta$, as in~\eqref{in-in}.
The perturbations see an interaction Hamiltonian $\Hint$,
of which we will consider the part cubic in the perturbations,
with time dependent coefficients due to the evolution of the background fields.
The perturbations are assumed to be initially in the Bunch-Davies vacuum,
but the non-linear
evolution introduces correlations between the modes.
As the modes cross the horizon they begin to behave classically
and eventually freeze out.

There is some freedom in how to represent the interaction Hamiltonian,
as the equation of motion of the free fields can be used, along with integration by parts~\cite{rp_integ_by_parts}.
This can be used, as pointed out in~\cite{Funakoshi}, to avoid numerically difficult cancellations.
Some presentations of this calculation use a field redefinition to eliminate terms
proportional to the equation of motion from the Lagrangian.
As pointed out in~\cite{px_burrage},
this is unnecessary as these terms will never contribute to the bispectrum result.
In fact, in some scenarios (such as resonant models) it introduces a numerically difficult
late time cancellation between a term in the interaction Hamiltonian and the
correction to the correlator that adjusts for the field redefinition.

The bispectrum arising from a single field inflation model,
with a canonical kinetic term, slowly rolling, turns out to produce
unobservably small non-Gaussianity~\cite{Maldacena}.
However, by breaking these assumptions large signals can arise.
These signals are usually calculated using~\eqref{in-in} within tailored approximations.
The results are not always separable, so further approximations must then be made to
allow comparison with the CMB.

\subsection{\boldmath $P(X,\phi)$ theories and approximate bispectrum templates}\label{sec:rev_px}
There is an extensive literature on the calculation
of bispectra from models of inflation~\cite{chen_easther_lim_1,chen_easther_lim_2,chen_ng_0605,seery_ng_0503,px_burrage,adshead,flauger_pajer_resonant,features_bartolo,bdy_passaglia}.
Multi-field models can produce large
correlations between modes of very different scales;
non-canonical kinetic terms can reduce the sound speed of the perturbations,
boosting both the smooth non-Gaussian correlations, and any
features which may be present~\cite{dbi_adshead,dbi_in_the_sky,dbi_miranda,dbi_silverstein,dbi_step_miranda,chen_folded_resonant,osc_avila};
effectively single-field models with imaginary sound speeds can generate a bispectrum
mostly orthogonal to the usual equilateral and local templates~\cite{RP_1}.
The methods outlined in this paper have been implemented
and tested for single-field models,
with multi-field models being a prime target for future work.
We will work with an inflaton action of the form
\begin{align}
S = \int d^4x \sqrt{-g}P(X,\phi)
\end{align}
with $X=-\frac{1}{2}g^{ab}\nabla_a \phi\nabla_b \phi$.
We work with the number of e-folds, $N$, as our time variable:
$x'=\frac{dx}{dN}=a\frac{dx}{da}$.
We define the Hubble parameter and the standard ``slow-roll'' parameters:
\begin{equation}
\label{slowrollparams}
\begin{split}
    H = \frac{d\ln a}{dt}	\,,
    \qquad
    \eps &= -\frac{d\ln H}{dN}	\\
    \eta = \frac{d\ln \varepsilon}{dN}	\,,
    \qquad
    \eps_s &= +\frac{d\ln c_s}{dN}	\,.
\end{split}
\end{equation}
though we make no assumption that these are actually small.
$c_s$ is the sound speed of the theory, which can vary with time:
\begin{align}
c_s=\frac{P,_X}{P,_X+2XP,_{XX}}.
\end{align}
The background quantities are evolved according to the Friedmann equations,
which are set with consistent initial conditions.
The equation of motion for the perturbations is:
\begin{align}\label{modefneqn}
\zeta_k''+(3-\varepsilon+\eta-2\varepsilon_s)\zeta_k'+\frac{c_s^2k^2}{a^2H^2}\zeta_k=0
\end{align}
where $c_s=1$ for standard canonical inflation.
We use standard Bunch-Davies initial conditions,
which leads us to impose the following condition deep in the horizon:
\begin{align}\label{bd_ic}
\zeta_k = \frac{i}{a}\sqrt{\frac{c_s}{4\varepsilon k}} e^{-ik\tau_s}
\end{align}
where we define $\tau_s$ through $\tau_s'=\frac{c_s}{aH}$
in analogy with the usual $\tau$ with $\tau'=\frac{1}{aH}$.
The solution in slow-roll (without features) is then approximately
\begin{align}\label{modefnsapprox}
    \zeta_k \propto (1+ik\tau_s)e^{-ik\tau_s}.
\end{align}
At leading order in slow-roll the power spectrum is~\cite{mukhanov_1999,chen_ng_0605}:
\begin{align}
P^{\zeta}(k) = \frac{1}{8\pi^2}\frac{H^2}{c_s\varepsilon},
\end{align}
where the right hand side is evaluated at $c_{s}k=aH$.
The spectral index is (also to leading order):
\begin{align}
n_s-1 = -2\varepsilon-\eta-\varepsilon_s.
\end{align}
Similarly to~\cite{Funakoshi}, at early times we extract the factor of $e^{-ik\tau_s}$ from the mode functions
and numerically evolve $\zeta_ke^{ik\tau_s}$\footnote{
    In fact~\cite{Funakoshi} extracts a factor of $e^{-ikc_s(\tau)\tau}$, losing efficiency
    due to slow-roll corrections.
}.
Unless interrupted, this prefactor decays exponentially.
Eventually we switch to evolving $\zeta_k$ directly.
For featureless slow-roll inflation the timing of the switch is simple;
so long as it is around horizon crossing, or a couple of e-folds after,
the precise location will not affect the result.
This becomes trickier when we are dealing with a model with
a step feature, for example.
Here, we found that navigating the feature in the first set of variables
causes difficulty for the stepper.
Switching to $\zeta_k$ before the onset of the feature
gives robust results without needing to loosen the tolerance.

Initially we consider the same basic models as in~\cite{Funakoshi};
a quadratic potential
\begin{align}\label{eq:quadratic_potential}
    \quadpot = \frac{1}{2}m^2\phi^2.
\end{align}
with a canonical kinetic term,
and a non-canonical model, with a DBI
kinetic term
\begin{align}\label{eq:dbi_action}
    S_{DBI}=\int d^4x\sqrt{-g}\left(-\frac{1}{f(\phi)}\left(\left(1+f(\phi)\partial_\mu\phi\partial^\mu\phi\right)^{\frac{1}{2}}-1\right)-V(\phi)\right),
\end{align}
with
\begin{align}\label{eq:dbi_warp}
    f(\phi)=\frac{\lambda_{DBI}}{\phi^4},\qquad
    V(\phi)=V_0-\frac{1}{2}m^2\phi^2,\qquad
    m=\sqrt{\beta_{IR}}H.
\end{align}

For our more stringent validation tests we work with feature model scenarios
based on the above base models.
To explore non-Gaussianity coming from sharp features we include
a kink
\begin{align}\label{eq:kink_potential}
    V(\phi) = \quadpot\left(1-c\tanh\left(\frac{\phi_f-\phi}{d}\right)\right).
\end{align}
To explore non-Gaussianity from deeper in the horizon we imprint
extended resonant features on the basic potential
\begin{align}\label{eq:resonant_potential}
    V(\phi) = \quadpot\left(1+bf\sin\left(\frac{\phi}{f}\right)\right).
\end{align}
For more details on these models, see~\cite{chen_easther_lim_2}.
To express the bispectrum results more compactly we use the symmetric polynomial notation employed in~\cite{FergShell_2}:
\begin{align}\label{shape_notation}
\begin{split}
    K_p &= \sum_{i=1,2,3} k_i^p, \\
    K_{pq} &= \frac{1}{\Delta_{pq}}\sum_{i\neq j} k_i^p k_j^q,   \\
    K_{prs} &= \frac{1}{\Delta_{prs}}\sum_{i\neq j\neq l} k_i^p k_j^r k_l^s,
\end{split}
\end{align}
where $\Delta_{pq}$ is $2$ if $p=q$, $1$ otherwise
and $\Delta_{prs}$ is $6$ if $p=r=s$, $2$ if $p=r\neq s$ (and permutations),
and $1$ if $p,r,s$ are all distinct.
With a canonical kinetic term, the slow-roll result for the shape is:
\begin{align}\label{malda_shape}
    S^{Malda}(k_1,k_2,k_3) &= A^{Malda} \left( (3\varepsilon-2\eta)\frac{K_3}{K_{111}}+\varepsilon \left(K_{12}+8\frac{K_{22}}{K}\right) \right),\\
    A^{Malda} &= -\frac{1}{32}\frac{H^4}{12\varepsilon^2}.
\end{align}
with $\eta=2\varepsilon$ for~\eqref{eq:quadratic_potential}.
At the primordial level, this is well approximated by the separable local template
\begin{align}\label{local_shape}
S^{local}(k_1,k_2,k_3) = \frac{k_1^2}{k_2k_3}+\frac{k_2^2}{k_3k_1}+\frac{k_3^2}{k_1k_2} = \frac{K_3}{6\,K_{111}}.
\end{align}
However, the amplitude of this shape is expected to be tiny,
and the dominant contributions (the squeezed configurations) are expected
to have no observable effect~\cite{Cabass_2016}.
The local template is in fact used to test for multi-field effects~\cite{Planck_NG_2015}.
For the featureless DBI scenario, the shape function is~\cite{dbi_in_the_sky}:
\begin{align}\label{dbi_shape}
    S^{DBI}(k_1,k_2,k_3) &= A^{DBI}\frac{K_5+2K_{14}-3K_{23}+2K_{113}-8K_{122}}{K_{111}K^2},\\
    A^{DBI} &= -\frac{1}{32}\frac{H^4}{12\varepsilon^2}\left(\frac{1}{c_s^2}-1\right),
\end{align}
to leading order in slow-roll.
Any constraint on the magnitude $A^{DBI}$ can be translated into one 
on the effective sound speed which from \textit{Planck} has a lower limit $c_s^{DBI} \geq 0.087$
at $95\%$ significance~\cite{Planck_NG_2015}.
The shape~\eqref{dbi_shape} can be approximated by the separable equilateral template
\begin{align}\label{equil_shape}
    S^{equil}(k_1,k_2,k_3) = \frac{(k_2+k_3-k_1)(k_3+k_1-k_2)(k_1+k_2-k_3)}{k_1k_2k_3}.
\end{align}
These templates can be modified to be more physically realistic by including
scaling consistent with the spectral index $n_s$~\cite{Planck_NG_2015}.
For example, we can add some scale dependence to the DBI model in a reasonable first approximation by including a prefactor
\begin{align}\label{dbi_ns_shape}
    S^{DBI-n_s}(k_1,k_2,k_3) &= {\left(\frac{k_1k_2k_3}{k^3_\star}\right)}^{n_s-1}S^{DBI}(k_1,k_2,k_3).
\end{align}
We now turn to feature templates.
The result of adding a feature of the form~\eqref{eq:kink_potential}
is to add oscillatory features of the form
\begin{align}\label{cos_shape}
    S^{\cos}(k_1,k_2,k_3) = \cos(w(k_1+k_2+k_3))
\end{align}
though more realistically there is some phase, shape dependence and a modulating envelope,
as detailed in~\cite{adshead}.
The result of adding a resonant feature of the form~\eqref{eq:resonant_potential}
is to generate logarithmic oscillatory features in the shape function of the form
\begin{align}\label{ln_cos_shape}
    S^{\ln-\cos}(k_1,k_2,k_3) = \cos(w\ln(k_1+k_2+k_3)).
\end{align}
With a non-canonical kinetic term, this can also
cause out-of-phase oscillations in the folded limit as well as a modulating shape,
see~\cite{chen_folded_resonant}.

Much success has been had in constraining non-Gaussianity
in the CMB using separable approximations to these approximate templates.
Other methods target oscillations~\cite{reso_estimator}, by expanding the shape function
in $k_1+k_2+k_3$, thus limiting their ability to capture shapes whose
phase varies across the tetrapyd.
Our motivation in this work for directly calculating the primordial
bispectrum in a separable form is to build towards
a pipeline to constrain a broader section of the model space,
removing these layers of approximations,
though these standard results provide useful validation tests.

\subsection{Configuration-by-configuration codes}\label{sec:rev_config_codes}
Previous work on the numerical calculations of inflationary
non-Gaussianity include the BINGO code~\cite{BINGO},
Chen et al~\cite{chen_easther_lim_1,chen_easther_lim_2},
the work of Horner et al~\cite{horner_methods,horner_ng,horner_cs}
and the Transport Method~\cite{transport_main,transport_pytransport,transport_pytransport_2,transport_curved_3_point}.
All but the last directly apply the tree-level in-in formalism $k$-configuration by $k$-configuration for a given model;
they integrate a product of three mode functions and a background-dependent term from the interaction Hamiltonian, of form similar to~\eqref{inin_sep}.
The eventual result is a grid of points representing the primordial bispectrum.

The most advanced publicly released code for the calculation of inflationary perturbations
is based on the Transport Method.
Like the previously mentioned work it calculates the bispectrum $k$-configuration by $k$-configuration.
However the method is different in its details.
Instead of performing integrals,
a set of coupled ODEs is set up and solved.
The power spectra and bispectra themselves are evolved, their time derivatives calculated by differentiating the in-in formalism.\footnote{
    One could imagine applying the same philosophy to our method.
    Certainly, at first sight this seems more natural, that if the core
    quantities in our method are the coefficients in some basis expansion,
    why not evolve them directly? Why take the apparently circuitous route
    of evolving the $\zeta_k(\tau)$, and decomposing them at every timestep?
    The answer is that the ``equations of motion'' for the coefficients of the expansion
    obtained by substituting the mode expansion of $\zeta_k(\tau)$ into~\eqref{modefneqn}
    are coupled in an infinite hierarchy.
}
The publicly released code is very sophisticated,
able to deal with multiple fields in curved field spaces,
recently being used to explore the bispectra resulting from
sidetracked inflation~\cite{RP_1}.

However despite the differences, all configuration-by-configuration methods face the same problems:
firstly, that calculating enough points in the bispectrum to ensure that
the whole picture has been captured is expensive, especially for non-trivial features.
Even once that has been achieved, what is obtained is a grid of points
which must be processed further to be usefully compared to observation.
Secondly, they must carefully implement some variation
of the $i\eps$ prescription without affecting the numerical results.
In~\cite{transport_main} this is achieved in the initial conditions for the bispectra;
other methods impose some non-trivial cutoff at early times.

\subsection{In-in separability}\label{sec:rev_funakoshi}
In~\cite{Funakoshi} it was pointed out that one can compute using the
tree-level in-in formalism in such a way as to preserve its intrinsic
separability. In addition to making this point,~\cite{Funakoshi} lays
out some of the basic structure of an implementation of that computation,
and validates the method on simple, featureless scenarios.
This work built on the philosophy of~\cite{FergShell_1,FergShell_2,FergShell_3}
in which a formalism was developed to
leverage the tractability of separable CMB bispectrum estimation
for generic primordial bispectra, by expanding them in a separable basis.
The results of these methods (not using the work of~\cite{Funakoshi})
are constraints on the parameters of certain inflation models through approximate
phenomenological templates.
These constraints can be found in~\cite{Planck_NG_2015, Planck_NG_2018}.
The idea of~\cite{Funakoshi} is an extension of that philosophy to the primordial level,
and our work is in implementing that idea.
In~\cite{FergShell_1,FergShell_2,FergShell_3} an orthogonal basis on the tetrapyd was used,
removing the need to fit non-physical configurations.
One of the main differences between that work and this
is that we cannot use this basis here without sacrificing the
in-in separability we are trying to preserve.

In this work we explore the details of this calculation in much greater detail
than was considered in~\cite{Funakoshi}.
We restructure the methods, improving on the work of~\cite{Funakoshi} in terms
of flexibility of basis choice and efficiency of the calculation.
We also detail a particular set of basis functions that improves upon those described
in~\cite{Funakoshi} in its rate of convergence, its transparency,
and its flexibility.
We do this without sacrificing orthogonality.
This is detailed in section~\ref{sec:choice_of_basis}.
Our improvements over the methods sketched in~\cite{Funakoshi} allow us to validate
on non-trivial bispectra for the first time, including sharp deviations from slow-roll, which we present in
section~\ref{sec:validation}.
We quote our results in terms of a measure that is
easier to interpret than the correlation defined in~\cite{Funakoshi},
and that includes the magnitude as well as the shape information
on the full tetrapyd.
This is discussed in section~\ref{sec:rev_precision_tests}.

\subsection{Precision tests}\label{sec:rev_precision_tests}
The inner product of two bispectrum shape functions is given by 
\begin{align}
    S_1\cdot S_2  = \langle S_1\,, S_2 \rangle = \int_{T_k} d^3k \: S_1(k_1,k_2,k_3) \: S_2(k_1,k_2,k_3)\,,\label{inner_prod}
\end{align}
where $T_k$ refers to the tetrapyd, the region of the cube $[\kmin,\kmax]^3$ that obeys the triangle inequality.
Following~\cite{hung_1902}
we define the two correlators:
\begin{align}
    \shapecor(S_1,S_2) = \frac{S_1\cdot S_2}{\sqrt{(S_1\cdot S_1)(S_2\cdot S_2})}\,, \qquad\quad 
\ampcor(S_1,S_2) = \sqrt{\frac{S_1\cdot S_1}{S_2\cdot S_2}}\,.
\end{align}
Here, we refer to $\shapecor(S_1,S_2)$ as the shape correlator between the two bispectra;
$\ampcor(S_1,S_2)$ is the amplitude correlator.
In principle, we could add some observationally motivated weighting
to the above measure, as considered in~\cite{FergShell_1,FergShell_2,FergShell_3},
but in this work we restrict ourselves to accurately calculating the
full primordial bispectra, weighting each configuration equally.

Writing $|S|^2=S\cdot S$,
we can then re-express a measure of the relative error
between one bispectrum template and another:
\begin{align}\label{relative_difference}
\totalcor(S_1,S_2) &= \sqrt{\frac{|S_1-S_2|^2}{|S_2|^2}}  = \sqrt{\frac{|S_1|^2-2S_1\cdot S_2+|S_2|^2}{|S_2|^2}}\nonumber\\
	   &= \sqrt{\ampcor(S_1,S_2)^2-2\ampcor(S_1,S_2)\shapecor(S_1,S_2)+1}.
\end{align}
This error measure takes into account differences in overall magnitude as well as shape.
If we are only interested in comparing the differences coming from the shape,
we can scale the bispectra so that $\ampcor(S_1,S_2)=1$ and so
\begin{align}\label{relative_difference_scaled}
    \totalcor(S_1,S_2) = \sqrt{2(1-\shapecor(S_1,S_2))}.
\end{align}
With this measure of relative difference, a shape correlation of $0.9$ corresponds to an error of $45\%$,
a shape correlation of $0.99$ corresponds to an relative difference of $14\%$,
a shape correlation of $0.999$ corresponds to an relative difference of $4\%$.
Thus this more exacting measure $\totalcor$ from~\cite{hung_1902} is a far better representation of actual convergence  between two shape functions
than the correlation used in~\cite{Funakoshi}.
We will use this measure to test the accuracy and efficiency of our basis expansion
in reconstructing the standard templates, and later to quantify the convergence
of our validation examples in section~\ref{sec:validation}.
In that section we also plot residuals on slices through the tetrapyd,
relative to the representative value
\begin{align}\label{rep_val}
    \sqrt{\frac{S\cdot S}{\int_{T_k} d^3k}}.
\end{align}

The squeezed limit of canonical single-field bispectra will not cause
observable deviations from a Gaussian universe,
due to a cancellation when switching to physical coordinates~\cite{Cabass_2016}.
Here, we will only consider primordial phenomenology
in comoving coordinates, so despite this cancellation,
the squeezed limit is still a useful validation test of our results,
using the standard single-field squeezed limit consistency condition~\cite{sqz_consistency,not_so_sqz}.
With $\mathbf{k_S}\equiv\left(\mathbf{k_2}-\mathbf{k_3}\right)/2 $:
\begin{align}\label{eq:sqz_consistency}
    S(k_1,k_2,k_3) = -\left[(n_s-1)|_{k_S}+\mathcal{O}\left(\frac{k_1^2}{k_S^2}\right)\right]P_{\zeta}(k_1)P_{\zeta}(k_S),
\ \ \  k_1\ll k_S
\end{align}
where $S(k_1,k_2,k_3)$ is again our dimensionless shape function.
That the error in the consistency relation decreases at least quadratically
in the long mode was shown in~\cite{not_so_sqz}.

\section{Methodology}\label{sec:methods}
Given its separable form, the tree-level in-in formalism is amenable
to more efficient calculation using separable modes, as first mentioned in~\cite{Funakoshi}.
That work extended the separable methodology previously implemented for the CMB bispectrum~\cite{FergShell_3}.
Our goal in this work is the efficient calculation of more general bispectra
which may have significant (possibly oscillatory) features, requiring searches across free parameter dependencies.
To achieve this, we represent the shape function~\eqref{shapefn} using a set of basis functions as
\begin{align}\label{goal}
S(k_1, k_2,k_3) &= \sum_n \alpha_n  \, Q_n(k_1,k_2,k_3)\,,
\end{align}
where the basis functions $Q_n(k_1,k_2,k_3)$ are explicitly separable functions of their arguments.
Translating this result into a constraint from the CMB
will require a large once-off computational cost, paid once
per set of basis functions $Q_n$,
not per scenario (encoded in $\alpha_n$).
The details of this once-per-basis calculation will be
presented in~\cite{Sohn_2020}.
As such, while the general computational steps we
describe will be independent of the basis, it is vital we
explore possible sets of basis functions $Q_n(k_1,k_2,k_3)$
and their effects on convergence;
we do this in section~\ref{sec:choice_of_basis}.
In section~\ref{sec:setting_notation} we set the notation we will use to recast
the standard numerical in-in calculation into a calculation of $\alpha_n$,
and sketch the steps involved.
In section~\ref{sec:h_int} we outline the details of the interaction Hamiltonian,
including accounting for the spatial derivatives in our final result.
In section~\ref{sec:k_dep} we make precise the numerical considerations
of the calculation,
especially our methods of dealing with the high-frequency
oscillations at early times.
\subsection{Choice of basis}\label{sec:choice_of_basis}
We begin our methods discussion by exploring possible sets of separable
basis functions $Q_n(k_1,k_2,k_3)$
for use in the expansion~\eqref{goal}.
Whether the goal is to explore primordial phenomenology or for direct comparison with observations,
the convergence of our basis set will determine the efficiency and practicality
of our methods.
We shall consider constructing the separable basis functions $Q_n(k_1,k_2,k_3)$
out of symmetrised triplet products of normalized one-dimensional modes $q_p(k)$ as
\begin{align}\label{modes3d}
    Q_n(k_1,k_2,k_3) \;\equiv\; {\Xi_{prs} } \, q_{(p} (k_1) \, q_{r}(k_2)\, q_{s)}(k_3)\,.
\end{align}
Here,  $n$ labels the ordered integer triplet $n \leftrightarrow \{p r s\}$ in an appropriate manner (see some ordering alternatives in~\cite{FergShell_3}), while the symmetrised average of all $\{p r s\}$ permutations is 
\begin{align}\label{modes3d}
    q_{(p}q_{r}q_{s)}\equiv(1/3!)\sum_\text{perms}q_{p}q_{r}q_{s}\qquad \text{and} \qquad 
     \Xi_{prs}= 
     \begin{cases}
     1, & p = r = s ~~~\text{all equal},\\
     \sqrt{3} & \{p r s\} ~~\text{any two equal,} \\
     \sqrt{6} & \{p r s\}  ~~\text{all  different.}
     \end{cases}
\end{align}
Unless stated otherwise, the $\{prs\}$ triples for each permutation set $n$
in~\eqref{modes3d} are represented by the coefficient with $0\le p\le r\le s$,
that is, $\alpha_n = \alpha_{prs} \equiv \alpha_{(prs)}$.
This modal expansion is terminated at some $\Pmax$ for which $\text{max}(p,r,s)<\Pmax$. 
Given the basis-agnostic methods we shall outline in the following 
sections~\ref{sec:setting_notation},~\ref{sec:h_int} and~\ref{sec:k_dep},
we are free to choose our set of basis functions to optimise for efficient convergence,
ensuring our results are useful for comparison with observations. 
There are a wide variety of options available, such as polynomial bases
or Fourier series, that can be chosen for the $q_p(k)$.
While not strictly necessary for the method, it is more convenient if 
the resulting 3D basis functions $Q_n(k_1,k_2,k_3)$
are orthogonal on the cubic region of selected wavenumbers, making it much more straightforward to obtain controlled convergence. Overall, then, rapid convergence is the key criterion in choosing the basis functions $q_p(k)$ in~\eqref{modes3d}, thus determining the nature of the numerical errors in the calculated bispectrum.
However, since we are going beyond the featureless examples of~\cite{Funakoshi}
this matter deserves considerable care and close attention.
Ideally we would have a three-dimensional basis that can efficiently
capture a wide variety of shapes on the tetrapyd, with relatively few modes.
In this work we aim for basis functions that work well in a wide variety of scenarios,
so we endeavour to use as little specific information as possible
(e.g.\ guessing the frequency of bispectrum oscillations from
the power spectrum of a given scenario), though we
will allow ourselves to use a representative value
of the scalar spectral index, $n_s^{*}$.   It is worth emphasising that a major advantage of the flexibility of the basis
in the methods detailed in the following sections is the ease with which
the basis can be modified to yield drastic increases in 
the rate of convergence at the primordial level, for the purposes of
exploring primordial phenomenology.

In this section we will use some standard templates to investigate different possible sets of basis functions.
An important issue is that when leveraging the separability of the in-in formalism,
we are essentially forced to expand the shape function on the entire cube ${[\kmin,\kmax]}^3$.
This is because the only decomposition we actually perform is
a one-dimensional integral over $[\kmin,\kmax]$
(as we will see in~\eqref{mode1Dcoeffs_integral}).
With a uniform weighting, this integral does not know anything about the 
distinction between the tetrapyd and the cube.
This is important as it means the non-physical configurations outside the tetrapyd will affect the convergence of our result on the tetrapyd, the region where we require efficient convergence.
To mimic this in testing our sets of basis functions, each shape will be decomposed on the
entire cube, but the quoted measures of convergence will be between the shape
and its reconstruction on the tetrapyd only (unless stated otherwise).

For a shape like~\eqref{malda_shape} the non-physical off-tetrapyd configurations
will not have a large effect, as the bispectrum on the faces of the cube is comparable
to the bispectrum in the squeezed limit of the tetrapyd.
On the other hand, for a shape of the equilateral type such as~\eqref{dbi_shape},
this effect can be disastrous if not handled properly.
This can be easily seen from~\eqref{equil_shape}, in the limit of small $k_3$.
The triangle condition in that limit enforces ${(k_2-k_1)^2\leq k^2_3}$.
This implies that ${0\leq k^2_3-(k_2-k_1)^2\leq k^2_3}$,
forcing the shape to go to zero in that limit despite the $k_3$
in the denominator.
On the non-physical part of the face, $k_2-k_1$ is not small,
and so the shape is boosted by $1/k_3$ relative to the equilateral configurations.
These regions then dominate any attempted basis expansion.
To overcome this problem, as we shall discuss, we will extend our basis to explicitly include this $1/k$ behaviour\footnote{
    There are results in the literature that describe
    generic $K=k_1+k_2+k_3$ poles
    in correlators---see for example~\cite{cosmo_bootstrap}.
    A simple example can be understood by recalling that
    in standard calculations using the in-in formalism, the $i\varepsilon$ prescription
    is used to damp out contributions in the infinite past.
    This does not work for $K=0$. While the resulting
    divergence (in $K$) is clearly outside the physical region of the tetrapyd,
    we will see its effects in the physical configurations.
    Given that this three-dimensional behaviour is generic, one might worry that we should
    take more care in building it into our one-dimensional basis.
    However, the excellent convergence in
    section~\ref{sec:validation}
    shows that $\Linvk$ and $\Lnsboth$ can capture this behaviour well,
    and that this worry is unwarranted.
    In fact, since this behaviour comes from the oscillations
    at early times, observing this behaviour is a useful
    check on our results.% in section~\ref{sec:validation}.
}.

\begin{figure}[!pth]
\centering     %%% not \center
\subfloat[Reconstructing the Maldacena Template]{\label{fig:a}\includegraphics[width=.49\columnwidth]{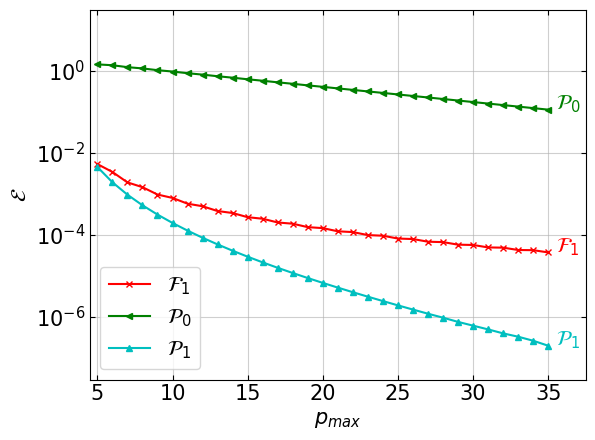}}
\subfloat[Reconstructing the DBI Template]{\label{fig:b}\includegraphics[width=.49\columnwidth]{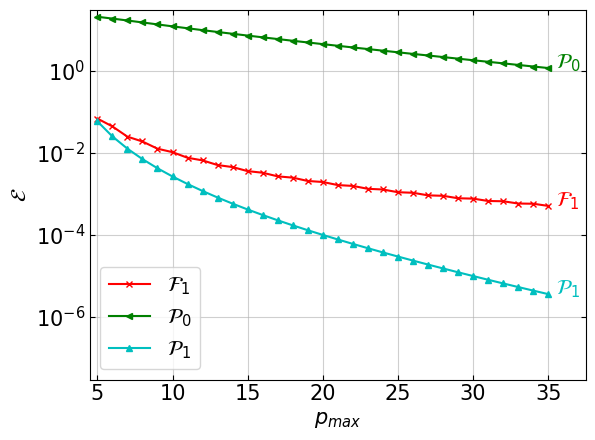}}
\caption{
    Convergence comparisons for the Legendre and Fourier basis functions for (a) the Maldacena 
    template~\eqref{malda_shape} and (b) the DBI template~\eqref{dbi_shape}.
    The pure Legendre $\Lbasic$ basis requires many terms to fit the $1/k$ behaviour
    in both Maldacena's template~\eqref{malda_shape} and the DBI template~\eqref{dbi_shape}.
    In contrast, the $\Linvk$ basis (with an orthogonalised $1/k$ term) mitigates this dramatically, with the 
    error already reduced by a factor of $100$ at $\Pmax=5$.
    The Fourier $\Finvk$ basis performs well, but converges more slowly than the $\Linvk$ basis.
    Note that the convergence errors for~\eqref{dbi_shape} are larger than~\eqref{malda_shape} because of the larger contributions outside the tetrapyd dominating the fit.
}\label{fig:recon_malda_dbi}
\end{figure}

To date the most useful starting choice for modal bispectrum expansions has been shifted Legendre polynomials $P_r(x)$:
\begin{align}\label{P0_definition}
    q_{r}(k) = \left(\frac{2r+1} {\kmax - \kmin}\right)^{1/2} P_r(\kbar)\,,\qquad 
\end{align}
with a rescaling of the argument $\kbar$ to ensure the wavenumber $k$ falls within the chosen (observable) domain $\kmin< k<\kmax$, that is, 
\begin{align}\label{krescaled}
    \kbar=\frac{2k-\kmax-\kmin}{\kmax-\kmin}\,.
\end{align}
We shall label as $\Lbasic$ the basis function set of pure Legendre polynomials in~\eqref{P0_definition},
with $r=0,1,\dots, p_{\rm max}\text{$-$}1$.
These were considered also in~\cite{Funakoshi}, however,
while they prove to be particularly functional building blocks for other modal applications,
in the context of the in-in formalism
they converge so slowly even for simple shapes as to be inadequate when taken on their own.
This poor rate of convergence with $\Lbasic$ for two local- and equilateral-type shapes is shown in figure~\ref{fig:recon_malda_dbi}.
It is due to the $1/k$ behaviour inherent in these shapes,
which is compounded in the equilateral models by pathologies exterior to the tetrapyd,
as we have discussed.
We can mitigate against this by including a basis function to capture this $1/k$ behaviour, 
\begin{align}\label{P1_definition}
    q_{p_{\rm max}}(k) = \text{Orth}\left[{1}/ {k}\right]\,, ~~ \quad \text{with} \quad  \Linvk = \{ q_r(k) ~\vert~ r= 0, 1,2, \dots , p_{\rm max}\}\,,
\end{align}
where $\text{Orth}$ represents the projection orthogonal to the original Legendre polynomial basis $\Lbasic$.
As we see in figure~\ref{fig:recon_malda_dbi}, the convergence properties for the augmented basis  $\Linvk$ are dramatically improved.

The two basis function sets actually used in~\cite{Funakoshi} to calculate primordial bispectra were as follows.
The first was the Legendre polynomials taken with a log-mapping
between $k$ and the polynomial argument as 
\begin{align}\label{Pln_definition}
    q_{r}(k) = \left(\frac{2r+1} {\ln\kmax - \ln\kmin}\right)^{1/2} P_r\left(\overline {\ln k}\right)\quad~ \text{with}\quad~ \overline {\ln k}=\frac{2\ln k-\ln\kmax-\ln\kmin}{\ln\kmax-\ln\kmin}\,.
\end{align}  
The second basis was implicitly mentioned in a reference to the possibility of multiplying the functions
to be decomposed by $k$, and dividing that factor out when evaluating
the result.
In our language, this is equivalent to working with an unnormalised 
basis set of
the Legendre polynomials divided by $k$:
\begin{align}\label{Pinv_definition}
        q_{r}(k) = \left(\frac{2r+1} {\kmax - \kmin}\right)^{1/2} \frac{P_r(\kbar)}{k}\,,
\end{align}
where the rescaled $\kbar$ is defined in \eqref{krescaled}.
This can also be thought of as expanding the bispectrum $k_1k_2k_3\,S(k_1,k_2,k_3)$ in $\Lbasic$,
instead of the shape function $S(k_1,k_2,k_3)$ itself. The consequence is that neither~\eqref{Pln_definition} nor~\eqref{Pinv_definition}
are orthogonal with respect to the flat weighting of the inner product~\eqref{inner_prod}.   However, as shown in~\cite{Funakoshi},
these two basis sets~\eqref{Pln_definition} and~\eqref{Pinv_definition}
are able to approximate the three canonical bispectrum shapes.   Nevertheless, our aim is to go beyond the featureless examples investigated in~\cite{Funakoshi},
so we require a basis that can capture many different forms of bispectrum features.
To this end, we prefer not to weight the large or small wavelengths in our fit,
as is done in~\eqref{Pln_definition} and~\eqref{Pinv_definition}.
The deciding factor for which weighting is optimal to include in the primordial inner product
is information about which configurations are most important for observables, that is, the expected signal-to-noise.
We will not discuss this matter in detail here, except to note that
motivated by the form of~\eqref{eq:reduced_cmb}, we will take as our aim the
accurate calculation of the primordial shape function with a flat weighting.
Based on this motivation, we will not pursue~\eqref{Pln_definition}
and~\eqref{Pinv_definition} any further.

One could certainly also consider sets of basis functions more tailored to a particular example,
or indeed even use power spectrum information to, on the fly,
generate a basis tailored to a rough form of the expected bispectrum features.
We save this possibility for future work.
In the following we will perform a more general exploration of orthogonal sets of basis functions that can
efficiently describe the necessary $1/k$ behaviour.
In addition to using the Legendre polynomials as building blocks,
we will also consider a Fourier basis for the purposes of comparison.

Our general strategy will be to augment these basic building blocks
with a small number of extra basis elements, while retaining orthogonality,
using the standard modified Gram-Schmidt process.
If we want to use some function $f$ to augment a given set of orthogonal functions $q_r$, with $r=0,\ldots,\Pmax$$-$$1$,
then we define
\begin{align}\label{gram_schmidt}
  \tilde {f}(k) \;=\;  \text{Orth}\left [f(k)\right] &~\equiv~ f(k) - \sum_{r=0}^{{\Pmax}\text{$-$}1}\frac{\left\langle f,q_r \right\rangle}{\left\langle q_r,q_r \right\rangle}q_r(k)
\end{align}
and add $\tilde{f}$ to our basis set, now of size $\Pmax+1$.
We note that the inner product here $\left\langle f,g \right\rangle$
is the 1D integral of the product $f(k)g(k)$ from $\kmin$ to $\kmax$.
The resulting basis is orthogonal, provided sufficient
care is taken to avoid numerical errors.

In addition to our Legendre basis functions, pure $\Lbasic$ and augmented $\Linvk$, we will also introduce a Fourier series basis denoted by $\Fbasic$ and defined by
\begin{align}\label{F0_definition}
    &q_{0}(k)    = 1\,, ~~
    q_{2r-1}(k)   = \sin(\pi r\kbar)\,,  ~~
    q_{2r}(k) = \cos(\pi r\kbar)\,,  ~~  1\leq r\leq(\Pmax\text{$-$}3)/2\\
    &q_{\Pmax\text{$-$}2}(k)=\bar k\,, ~~~q_{\Pmax\text{$-$}1}(k)=\bar k^2\,.
    \end{align}
Here, even the basic Fourier series have to be augmented by the linear $k$ and quadratic $k^2$ terms (for a total size of $\Pmax$),
in order to satisfactorily approximate equilateral shapes (reflecting in part the preference for periodic functions). 
As with $\Linvk$ defined in \eqref{P1_definition}, we will similarly create an augmented Fourier basis $\Finvk$ by adding the inverse $1/k$ term to the  $\Fbasic$ basis, i.e.\ using $q_{p_{\rm max}}(k) = \text{Orth}\left[{1}/ {k}\right]$ with \eqref{gram_schmidt}.
When we refer to convergence, we mean in increasing number
of Legendre polynomials (or sines and cosines) within the initial set.
The total size of the set will always be referred to as $\Pmax$.

\begin{table}[h!]
  \begin{center}
    \begin{tabular}{c|c|l|c}
      \textbf{Notation} & \textbf{Building Blocks} & \textbf{Augmented by} & \textbf{Definition}\\
      \hline
        $\Lbasic$  & Legendre polynomials &                & \eqref{P0_definition}\\
        $\Fbasic$  & Fourier Series & $k$, $k^2$           & \eqref{F0_definition}\\
        $\Linvk$   & Legendre polynomials & $k^{-1}$       & \\
        $\Finvk$   & Fourier Series & $k$, $k^2$, $k^{-1}$ & \\
        $\Lnsinv$  & Legendre polynomials & $k^{-1+(n_s^{*}-1)}$   & \eqref{Lns_inv}\\
        $\Lnsboth$ & Legendre polynomials& $k^{n_s^{*}-1}$, $k^{-1+(n_s^{*}-1)}$ & \eqref{Lns_both}\\
    \end{tabular}
    \caption{
          Basis summary---the augmentation of the basis is done
          using~\eqref{gram_schmidt}. The size of each basis is
          referred to as $\Pmax$.
      }\label{tab:basis_summary}
  \end{center}
\end{table}

In order to compare the efficacy of these four different basis function sets ($\Lbasic$, $\Fbasic$, $\Linvk$ and $\Finvk$), we have investigated their convergence on Maldacena's shape~\eqref{malda_shape} and the DBI shape~\eqref{dbi_shape}.  To mimic the in-in calculation, we expand the shape on the cube, but test the result on the tetrapyd
using~\eqref{relative_difference}.
The results are shown in figure~\ref{fig:recon_malda_dbi} where we find that the Legendre polynomials basis set $\Lbasic$ converges so slowly as to be unusable (with the Fourier modes $\Fbasic$ worse and not plotted).
However, the augmented Legendre basis $\Linvk$ (including $1/k$) leads to rapid convergence with an improvement of four orders
of magnitude at $\Pmax=15$. The augmented Fourier basis $\Finvk$ also converges
quickly relative to $\Lbasic$, but is outdone by $\Linvk$.
Though we do not show the convergence on the cube, we find that for Maldacena's
template this is of the same order of magnitude as
the error on the tetrapyd.  For the DBI shape, however, the
fit on the tetrapyd lags significantly behind, due to the effect
of the large non-physical configurations. This explains the order of
magnitude difference between the convergence at
each $\Pmax$ for the two shapes in figure~\ref{fig:recon_malda_dbi}.

\begin{figure}[!pth]
\centering     %%% not \center
\subfloat[$\cos(f(k_1+k_2+k_3))$]{\label{fig:a}\includegraphics[width=.49\columnwidth]{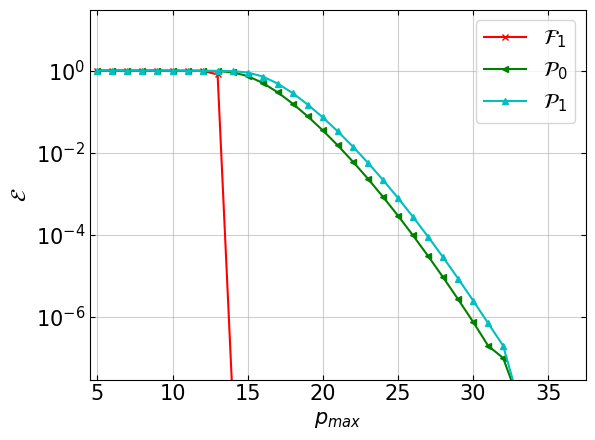}}
\subfloat[$\cos(f(k_1+k_2+k_3))S^{DBI}$]{\label{fig:a}\includegraphics[width=.49\columnwidth]{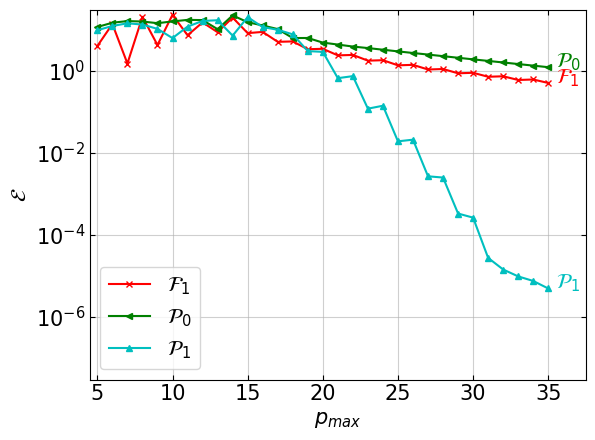}}
\caption{
    Convergence comparison for oscillatory models. (a) As expected, the $\Finvk$ basis fits an oscillation with
    no shape dependence~\eqref{cos_shape} (that is periodic in the $k$-range) perfectly.
    For this special case, the $\Lbasic$ and $\Linvk$ sets of basis functions require more modes
    to accurately describe the shape. (b) However, moving to the more complex and realistic case of a
    feature with scale and shape dependence (in this case the product of~\eqref{dbi_shape}
    and~\eqref{cos_shape}), we see that again $\Linvk$ converges with the fewest modes.
    Note that before the expansion has fully converged, the fit on the tetrapyd
    can actually degrade slightly when the basis set is extended. This is an artifact
    of fitting on the cube and restricting~\eqref{relative_difference}
    to the physical configurations on the tetrapyd; when considered over the
    entire cube the fit improves monotonically.
}\label{fig:recon_osc_dbiosc}
\end{figure}

\begin{figure}[!pth]
\centering     %%% not \center
\subfloat[$\cos(f\log(k_1+k_2+k_3))$]{\label{fig:a}\includegraphics[width=.49\columnwidth]{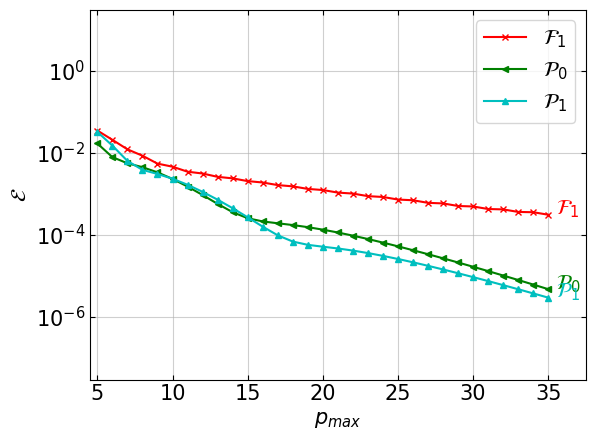}}
\subfloat[$\cos(f\log(k_1+k_2+k_3))S^{DBI}$]{\label{fig:a}\includegraphics[width=.49\columnwidth]{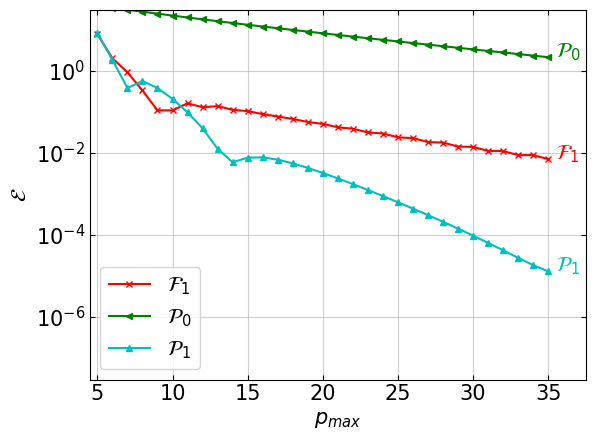}}
\caption{
    (a) The convergence for a log oscillation model~\eqref{ln_cos_shape} with no shape dependence.
    For this type of feature, the $\Lbasic$ and $\Linvk$ sets of basis functions require fewer modes
    to accurately describe the shape than the $\Finvk$ basis.
    (b) For the more complex and realistic case of a
    feature with scale and shape dependence (in this case the product of~\eqref{dbi_shape}
    and~\eqref{ln_cos_shape}), we see that again $\Linvk$ converges with the fewest modes,
    though in this case $\Finvk$ outperforms $\Lbasic$.
}\label{fig:log_recon_osc_dbiosc}
\end{figure}

Next, we investigate oscillatory model templates.
The simple feature model~\eqref{cos_shape} and the resonance model~\eqref{ln_cos_shape}
have scale dependence, but no shape dependence (in that they only depend on the perimeter
of the triangle, $K=k_1+k_2+k_3$).
We test our sets of basis functions on these two shapes,
and also when they are multiplied by~\eqref{dbi_shape} to obtain a feature
template with both shape and scale dependence.
As shown in figure~\ref{fig:recon_osc_dbiosc},
$\Fbasic$ naturally outperforms $\Lbasic$ for a pure oscillation,
but when the equilateral-type DBI template~\eqref{dbi_shape} is superimposed, even the augmented Fourier modes
$\Finvk$ converge poorly and the Legendre modes $\Linvk$ clearly offer a better more robust option.
In figure~\ref{fig:log_recon_osc_dbiosc} we see that for a logarithmic
oscillation, $\Linvk$ always converges in the fewest modes.


\begin{figure}[!pth]
\centering
\includegraphics[width=0.8\columnwidth]{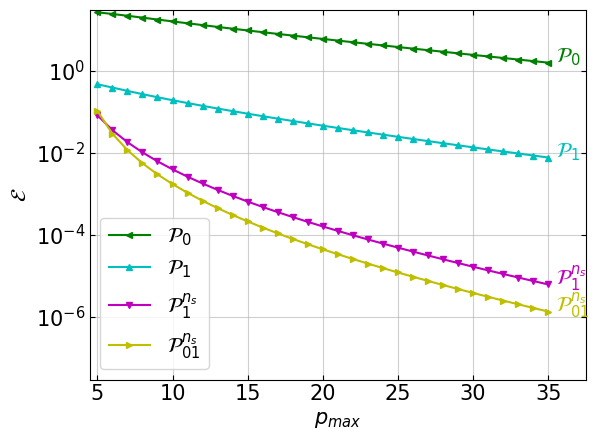}
\caption{
    For the scale-dependent DBI template~\eqref{dbi_ns_shape},
    by including a minimal amount of power spectrum information
    using~\eqref{Lns_inv} and~\eqref{Lns_both} (here with
    $n_s^{*}=0.97$), we can push the errors
    to less than $0.1\%$ at $\Pmax=15$,
    allowing us to work with a basis that can efficiently
    capture the expected scale dependence of the shape function.
    }\label{fig:recon_dbi_ns}
\end{figure}

Finally, we consider convergence in the light of the more subtle scale-dependence due to the spectral index $n_s$ of the power spectrum.  The simple canonical examples in figure~\ref{fig:recon_malda_dbi} had shape dependence and no scale dependence, but this would only be expected of scenarios unrealistically deep in the slow-roll limit.
When we include this scale dependence, using~\eqref{dbi_ns_shape} with $n_s$, 
it proves very useful to include these deviations from
integer power laws in the basis functions.  We consider two cases, first augmenting $\Lbasic$ by a scale-dependent $1/k$ term using the orthogonalisation procedure \eqref{gram_schmidt}, 
\begin{align}\label{Lns_inv}
    q_{\Pmax}(k) = \text{Orth}\left[{k^{-1+(n_s^{*}-1)}}\right],
\end{align}
which we refer to as $\Lnsinv$.
Secondly, we also augment $\Lbasic$ with an additional scale-dependent constant term as 
\begin{align}\label{Lns_both}
    q_{\Pmax}(k) = \text{Orth}\left[{k^{(n_s^{*}-1)}}\right],\qquad  q_{\Pmax\text{$+$}1}(k) = \text{Orth}\left[{k^{-1+(n_s^{*}-1)}}\right],
\end{align}
which we refer to as $\Lnsboth$.
As we see in figure~\ref{fig:recon_dbi_ns}, for equilateral type shapes
even a small overall scale dependence causes significant degradation in the convergence of the original augmented Legendre basis $\Linvk$.  However, incorporating the spectral index $n_s$  into the basis functions $\Lnsinv$ and $\Lnsboth$ results again in rapid convergence to the scale-dependent DBI template, which can be accurately approximated with a limited number of modes.    We conclude that augmenting the basis functions with terms incorporating the expected dependence on the spectral index enables the efficient approximation of high precision primordial bispectra. 

\subsection{Exploiting the separability of the in-in formalism}
\label{sec:setting_notation}
In this section we set up the notation, and sketch the steps required to calculate the
coefficients $\alpha_n$ in~\eqref{goal}.
The values of these coefficients will depend on the choice
of basis, but the description of the methods below will remain mostly basis agnostic.
Our aim will be to separate out the dependence on $k$ and $\tau_s$,
without losing information, except in the sense that is controlled by $\Pmax$.
We will set up an efficient numerical implementation of the calculation,
a necessary consideration to allow this method to be useful in exploring
parameter spaces in primordial phenomenology.
Throughout we will see that we are able to preserve the
separability of the dependence on $k_1$, $k_2$ and $k_3$.

The tree-level in-in formalism for the bispectrum~\eqref{in-in} is inherently separable given
the form of the cubic interaction Hamiltonian $H_{\rm int}$.
Consider indexing with ${i=1,2,3...}$ the interaction vertices in $H_{\rm int}$,
so then the bispectrum~\eqref{in-in} can be expressed as a sum over separable contributions of the form:

\begin{eqnarray}\label{inin_sep}
S(k_1, k_2,k_3) &=& \sum_i I^{(i)} (k_1, k_2,k_3)\nonumber \\
    &=& \sum_i \bigg[ v^{(i)}(k_1, k_2,k_3) \int^0_{-\infty(1-i\varepsilon)} d\tau\, w^{(i)}(\tau) F^{(i)}(\tau,k_1)\, G^{(i)}(\tau,k_2)\,J^{(i)}(\tau,k_3)\nonumber\\
    &&\quad\quad\quad\quad\quad\quad\quad\quad\quad\quad+ \text{cyclic perms}  \bigg] 
\end{eqnarray}
where $w^{(i)}(\tau)$ is a function of the scale factor and the other background parameters~\eqref{slowrollparams} for the $i$-th interaction vertex, while the terms $F^{(i)}, G^{(i)}, J^{(i)}$ are given by the Fourier mode functions $k^2\zeta_{\bf k}(0)\zeta^{*}_{\bf k}(\tau)$ or their time derivatives $k^2\zeta_{\bf k}(0)\zeta^{*'}_{\bf k}(\tau)$.
Spatial derivative terms, such as  $\partial_i \zeta\, \partial_i\zeta \longrightarrow ({\bf k_2}\cdot {\bf k_3} )\zeta_{{\bf k}_2}^*(\tau)\, \zeta_{{\bf k}_3}^*(\tau)$ also separate because of the triangle condition~\eqref{triangle_condition} giving  ${\bf k_2}\cdot {\bf k_3} = (k_1^2 - k_2^2-k_3^2)/2$, yielding a sum of separable terms.
These time-independent contributions are contained in $v^{(i)}(k_1, k_2,k_3)$,
as they do not force us to compute extra time integrals.
Note that $v^{(i)}(k_1, k_2,k_3)$ need not be symmetric in its arguments.

The terms contained in $v^{(i)}(k_1, k_2,k_3)$ depend on the structure of the spatial derivatives in the interaction Hamiltonian,
but not the specific scenario. These terms are separable; for details, see section~\ref{sec:h_int}.
We include their contribution to the final result after the time integrals have been computed.
The factors which depend only on time, $w_i(\tau)$, depend on the scenario
but do not need to be decomposed in $k$.
The remaining factors have both $k$ and time dependence;
they must be decomposed in $k$ at every timestep.
These terms look like $F^{(i)}(k, \tau) = k^{2} \zeta_{\bf k}(0)\zeta_{\bf k}^*(\tau)$
(or $k ^{2} \zeta_{\bf k}(0){\zeta_{\bf k}^*}'(\tau)$),
where $k^2$ comes from using the weighting of the scale-invariant shape function~\eqref{shapefn}.
This could be absorbed into $v^{(i)}(k_1, k_2,k_3)$, but we have the freedom to keep it here to aid convergence.

If the expressions being expanded have some known pathology in their $k$-dependence,
we can then see two ways of dealing with this.
The basis can be augmented to efficiently capture the relevant behaviour
(see section~\ref{sec:choice_of_basis})
or the behaviour can be absorbed into the analytic prefactor, $v^{(i)}(k_1, k_2,k_3)$.\footnote{At early times the modes are highly oscillatory in both $k$ and $\tau_s$, which certainly requires special attention.
We discuss this in section~\ref{sec:k_dep}.}
We use the former, as the numerics of the latter are less transparent
and less physically motivated.

The internal basis used for the decomposition at each timestep need not match that
which is used for the final result,
and indeed in dealing with the spatial derivatives in section~\ref{sec:h_int}
we will find it useful to change to a different basis than the one used to perform the
time integrals of the decompositions.

Using the approximate mode functions~\eqref{modefnsapprox}, an explicit example for the first interaction term in~\eqref{interaction_loc}, i.e.\ $H_{\rm int}^{(1)} = {\zeta'}^2\zeta$, takes the form
\begin{align}\label{inin_example}
    F^{(1)}(\tau,k) = G^{(1)}(\tau,k)  ~=~  c_s k^2 \tau\frac{H^2}{4\varepsilon c_s} e^{ic_{\rm s} k\tau} \, , \qquad\,
    J^{(1)}(\tau,k) ~=~ (1-ik c_{\rm s}\tau)\frac{H^2}{4\varepsilon c_s}e^{ic_{\rm s} k\tau} \,.
\end{align}
In the simple mode approximation~\eqref{modefnsapprox}, such terms in~\eqref{inin_sep} are straightforward to integrate analytically (using the $i\varepsilon$ prescription), provided the time-dependence of the slow-roll parameters and the sound speed is neglected~\cite{Maldacena}.   However, for high precision bispectrum predictions we must incorporate the full time-dependence, while solving~\eqref{modefneqn} to find accurate mode functions $\zeta_{\bf k}(\tau)$ numerically.  Obtaining the full 3D bispectrum directly is computationally demanding at high resolution because it requires repetitive integration of~\eqref{inin_sep} at each specific point for the wavenumbers $(k_1, k_2, k_3)$, a problem which is drastically compounded by bispectrum parameter searches e.g. for oscillatory models. 

Consider representing the primordial shape function $S(k_1, k_2, k_3)$ in~\eqref{inin_sep} as
a mode expansion for each interaction term $I^{(i)}(k_1, k_2,k_3)$ as
\begin{align}\label{modeexp}
S(k_1, k_2,k_3) &= \sum_i I^{(i)}(k_1,k_2,k_3) =  \sum_i \sum_n \alpha_n^{(i)}  \, Q_n(k_1,k_2,k_3)\,,
\end{align}
where $Q_n(k_1,k_2,k_3)$ is separable, built out of some
orthonormal set $q_p(k)$ as in~\eqref{modes3d}.
Armed with this set of modes,
we can expand any of the interaction terms $F^{(i)}(\tau, k)$, $G^{(i)}(\tau, k)$, $J^{(i)}(\tau, k)$
in~\eqref{inin_sep} as:
\begin{align}
    F^{(i)}(\tau, k) &= \sum _p f_p^{(i)} (\tau) \, q_p(k)\,,\label{mode1Dcoeffs_sum}\\
    \text{where} \quad f_p^{(i)} (\tau)  &= \int _\kmin^\kmax dk \,F^{(i)}(\tau, k) \, q_p(k) \,.\label{mode1Dcoeffs_integral}
\end{align}
Note that in the simple mode approximation, as in~\eqref{inin_example},
we must expand $e^{ic_{\rm s} k\tau}$ in the terms of the $q_p(k)$.
At early times $\tau$ is large, so in $k$ this is highly oscillatory.
This creates two problems. Firstly, this seems to require many samples
in $k$ to accurately calculate each $f_p^{(i)} (\tau)$,
adding more modes that must be evolved in time.
To bypass this, we extract the oscillatory part at
early times, reducing the number of needed
$k$-samples; see section~\eqref{sec:k_dep} for details.
Secondly, it forces us to calculate $f_p^{(i)} (\tau)$
up to very high $p$ if we want to accurately converge to $F^{(i)}(\tau, k)$,
for sets of basis functions such as the Legendre polynomials.
In fact, obtaining a convergent final bispectrum result
does not require calculating the full convergent sum
for $F^{(i)}(\tau, k)$ in~\eqref{mode1Dcoeffs_sum},
as the highly oscillatory parts will cancel in the time integrals
for any sufficiently smooth $S(k_1,k_2,k_3)$.

Substituting these expansions into~\eqref{inin_sep}, we obtain the following decomposition for the $i$-th vertex contribution,
\begin{eqnarray}\label{inin_kdep}
    &&I^{(i)}(k_1,k_2,k_3)\nonumber\\
    &=& v^{(i)}(k_1, k_2,k_3)\int d\tau\, w^{(i)}(\tau)\, \sum_p f_p^{(i)}(\tau) q_p(k_1)\sum_r g_r^{(i)}(\tau) q_r(k_2)\sum_s h_s^{(i)}(\tau) q_s(k_3) + \text{cyclic perms}\nonumber\\	
	&=& v^{(i)}(k_1, k_2,k_3)\sum_{prs}\left(\int d\tau\, w^{(i)}(\tau)\, f_p^{(i)}(\tau)\, g_r^{(i)}(\tau) \,h_s^{(i)}(\tau)\right)\threeqs + \text{cyclic perms}\nonumber\,.
\end{eqnarray}
For the sake of compactness we use $P$ to stand for the triplet $p,r,s$ and $\tilde{P}$ to stand for the triplet $\tilde{p},\tilde{r},\tilde{s}$.
Writing $q_{P}(k_1,k_2,k_3)=\threeqs$,
we continue,
\begin{eqnarray}\label{inin_kdep}
    I^{(i)}(k_1,k_2,k_3) & =& v^{(i)}(k_1, k_2,k_3) \sum_{P} \tilde{\alpha}_{P}^{(i)}\,  q_{P}(k_1,k_2,k_3) + \text{cyclic perms}\nonumber\\
    & =& \sum_{P} \tilde{\alpha}_{P}^{(i)}\sum_{\tilde{P}}V^{(i)}_{P\tilde{P}}\,  q_{\tilde{P}}(k_1,k_2,k_3) + \text{cyclic perms}\nonumber\\
    & =& \sum_{\tilde{P}} \alpha_{\tilde{P}}^{(i)}\,  q_{\tilde{P}}(k_1,k_2,k_3) + \text{cyclic perms} \,,
\end{eqnarray}
where we have written
\begin{eqnarray}\label{inin_kindep}
\tilde{\alpha}_P^{(i)} =  \tilde{\alpha}_{prs}^{(i)} 	= \int d\tau\, w^{(i)}(\tau)\, f_p^{(i)}(\tau) \,g_r^{(i)}(\tau) \,h_s^{(i)}(\tau)\,,
\end{eqnarray}
and included the time-independent $k$-prefactors from the interaction Hamiltonian by writing
\begin{eqnarray}\label{V_definition}
    v^{(i)}(k_1, k_2,k_3)q_P(k_1,k_2,k_3) &=& \sum_{\tilde{P}}V^{(i)}_{P\tilde{P}}q_{\tilde{P}}(k_1,k_2,k_3)\,,
\end{eqnarray}
and
\begin{eqnarray}
    \alpha_P^{(i)} = \sum_{\tilde{P}} \tilde{\alpha}_{\tilde{P}}^{(i)}V^{(i)}_{\tilde{P}P}.
\end{eqnarray}
We connect to the coefficients of the ordered, symmetrised basis in~\eqref{modeexp} by taking the symmetry factor into account,
\begin{eqnarray}
    \alpha_n^{(i)} = \frac{3!}{\Xi_P}\alpha_P^{(i)}.
\end{eqnarray}
The numerical calculation of $V^{(i)}_{P\tilde{P}}$ (as defined by~\eqref{V_definition}) is highly efficient as $v^{(i)}(k_1, k_2,k_3)$
is a sum of separable terms. The details of these terms
depend only on the spatial derivatives in the interaction Hamiltonian, not the scenario being
considered, so the matrix can be precomputed and stored.
Note that this is not the only way one can organise this calculation to
explicitly preserve the separability. One could also include the contributions
coming from the spatial derivatives first, decomposing (as in~\eqref{mode1Dcoeffs_sum}) not only terms
like $k^{2} \zeta_{\bf k}(0)\zeta_{\bf k}^*(\tau)$, but also terms that include
each power of $k_1$, $k_2$ or $k_3$ that appears in $v(k_1,k_2,k_3)$. The index $i$
in the sum in~\eqref{modeexp} would then run over not only each vertex in the interaction
Hamiltonian, but also each separable term within those vertices.
We do not choose this path as, for the sake of efficiency, we wish to minimise the
number of time integrals of the form~\eqref{inin_kindep} we need to calculate.

Note the basis sets on the left and right hand side of~\eqref{V_definition}
need not match. In fact, if those two basis sets do match, then generically information
will be lost---for example, if the basis set on the left is $\Lbasic$, then terms
in $v^{(i)}(k_1,k_2,k_3)$ with positive powers will introduce higher order
dependencies on $k$, and negative powers will introduce $1/k$ behaviour. In
practice, to prevent this loss of information, we take the basis set on the right hand
side of~\eqref{V_definition} to be an expanded version of that on the left. For
example, if the left hand basis was $\Lbasic$ of size $\Pmax$, the right hand basis would
be $\Linvk$ of size $\Pmax+3$.

We can see from~\eqref{inin_kindep} that the number of time integrals needed is controlled by
$N_V\times\Pmax^3$\footnote{In fact the number is not quite $\Pmax^3$.
Since we have extracted the spatial derivatives, the only remaining
possible source of asymmetric $k$-dependence
comes from $\zeta^3$, $\zeta^2\zeta'$, $\zeta'^2\zeta$ or $\zeta'^3$
so the time integral in~\eqref{inin_kindep} will always be (at least) symmetric in $p$ and $r$. },
where $N_v$ is the number of interaction vertices
and $\Pmax$ is the size of the final basis.
Since the calculational cost of doing the internal decompositions depends only
linearly on the size of internal basis, improvements there are dwarfed by improvements
gained from reducing the number of terms needed in the final basis.

We will calculate the contribution of each $\Hint$ vertex separately,
indexing the vertices as above by $(i)$, so the overall shape function~\eqref{inin_sep},~\eqref{modeexp} is then simply
\begin{eqnarray}\label{shapemodeexp}
    S(k_1, k_2,k_3) =   \sum_n \left(\sum_i \alpha_n^{(i)}\right) \, Q_n(k_1,k_2,k_3) =  \sum_n \alpha_n \, Q_n(k_1,k_2,k_3) \,.
\end{eqnarray}
Depending on the scenario,  some vertex contributions will converge faster than others 
or be completely negligible;
for efficiency the maximum modal resolution defined by $\Pmax$ can be allowed to be different
for each vertex.

The raison d'etre for this approach is that all time integrals~\eqref{inin_kindep} are now independent of the $k$-configuration\footnote{
    They are not independent of $\kmin$ and $\kmax$ which define the domain of interest,
    which is analogous to the coefficients of a Taylor expansion depending on its expansion point.
}.
In a configuration-by-configuration method one improves the precision by
decreasing the spacing which defines the density of the grid of points within the tetrapyd.
Instead, in the modal approach, we increase precision by adding more modes to the shape function expansion~\eqref{shapemodeexp}
until the result converges at high precision.   At first sight, this appears to increase the dimensionality of the calculation.
Directly integrating the in-in formalism requires one time
integration for each $k$-configuration, i.e. $N_k^3$ integrals, ignoring symmetry.
The method detailed here requires decomposing the modes,
then a time integral for every coefficient, i.e. $\Pmax^3$ integrals
(again ignoring symmetry) plus the decomposition.
However for every model we have explored from the literature,
our expansion in $\Pmax$ converges far faster than in the number of
$k$-modes that would be required to have confidence in a sampled bispectrum.
This is clear in smooth bispectra such as~\eqref{malda_shape} and~\eqref{dbi_shape},
but is also true of bispectra with complicated features,
as seen in section~\ref{sec:choice_of_basis}.

To be efficiently connected to a late-time observable
a sampled bispectrum would have to be fit by a smooth template,
a complication that is automatically taken care of in this formalism.
We note that while the primordial basis is chosen for computational speed and convenience, it can be 
independent of the final bispectrum basis employed for observational tests; a change of basis $Q_n\rightarrow \tilde Q_n$ can be achieved through a linear transformation $\Gamma$ with the new expansion coefficients given by $\tilde \alpha_m = \Gamma_{mn}\alpha_n$.

Discussion of convergence in this work is considered only at the primordial level,
with no concept of the signal to noise of an actual experiment.
There could be a basis that converges faster in some observationally weighted sense,
efficiently describing the primordial modes which will matter most at late times.
We leave discussion of this point to a later work, as converting between the two,
after the in-in computation is completed, is trivial.

Having now set our notation and outlined the calculation, in the following sections we
discuss the actual numerical implementation of these methods.

\subsection{The interaction Hamiltonian}\label{sec:h_int}
The methods detailed in the previous section depend on the separability of
the third-order interaction Hamiltonian, $\Hint$,
and the possibility of including the spatial derivatives in a
numerically accurate and efficient way.
To make precise how our methods take into account the details of
$\Hint$, we will take $P(X,\phi)$ inflation as an example.
The full cubic interaction Hamiltonian, not neglecting boundary terms,
can be calculated as~\cite{px_burrage,chen_ng_0605,seery_ng_0503}
\begin{align}
    \Hint(t)=\ \int d^3x\bigg\{& -\frac{a^3\varepsilon}{H c_s^4}\left(1-c_s^{2}-2c_s^{2}\frac{\lambda}{\Sigma}\right)\dot{\zeta}^3
		+ \frac{a^3\varepsilon}{c_s^{4}}\left(3-3c_s^2-\varepsilon+\eta\right) \zeta\dot{\zeta}^2\nonumber\\
		&- \frac{a\varepsilon}{c_s^{2}}\left(1-c_s^2+\varepsilon+\eta-2\varepsilon_s\right) \zeta(\partial\zeta)^2\nonumber\\
        &- \frac{a^3\varepsilon^2}{2c_s^{4}}(\varepsilon-4)\dot{\zeta}\partial\zeta\partial(\partial^{-2}\dot{\zeta})
        - \frac{a^3\varepsilon^3}{4c_s^4}\partial^2\zeta(\partial(\partial^{-2}\dot{\zeta}))^2\bigg\}\label{interaction_loc}
\end{align}
with $\Sigma=\frac{H^2\varepsilon}{c^2_s}$
and $\lambda = X^2P,_{XX}+\frac{2}{3}X^3P,_{XXX}$.
See~\cite{px_burrage} for further details.

This is commonly quoted with a term proportional to the equation of motion,
but this will never contribute~\cite{px_burrage,bdy_arroja,bdy_passaglia,bdy_rigopoulos}.
We do not need to make a slow-roll approximation
(the quantities defined in~\eqref{slowrollparams} are not required to be small,
except in that we wish to have a successful inflation scenario),
nor do we need to neglect any terms in the interaction Hamiltonian.
We do no field redefinition,
so do not need to add a correction to the final bispectrum.
Following the calculation of~\cite{px_burrage} (see also~\cite{bdy_arroja,bdy_passaglia,bdy_rigopoulos})
we do not work with any boundary terms.
Numerically this is preferable to forms with boundary terms,
whether they come from undoing a field redefinition or from integration by parts.
Since the boundary term contribution will depend on the choice of when to end the integration,
its time dependence must cancel with a late-time time-dependent
contribution of some vertex, requiring us to track the necessary quantities
much longer than otherwise needed to obtain the desired precision.

Schematically, the correction from a field redefinition would look like
\begin{align}
{
\left< \zeta_{\bf{k_1}}\zeta_{\bf{k_2}}\zeta_{\bf{k_3}} \right>
    = \left< \tilde{\zeta}_{\bf{k_1}}\tilde{\zeta}_{\bf{k_2}}\tilde{\zeta}_{\bf{k_3}} \right>
    + \lambda \left< \tilde{\zeta}_{\bf{k_1}}\tilde{\zeta}_{\bf{k_2}} \right>\left< \tilde{\zeta}_{\bf{k_1}}\tilde{\zeta}_{\bf{k_3}} \right>
    + cyclic
}
\end{align}
where $\lambda$ is some function of the slow-roll parameters.
The correction terms will have a time dependence from $\lambda$,
so the $\left< \tilde{\zeta}_{\bf{k_1}}\tilde{\zeta}_{\bf{k_2}}\tilde{\zeta}_{\bf{k_3}} \right>$
term must have some late time contribution to cancel it.
To obtain an accurate result, care would need to be taken with this cancellation,
an unnecessary complication.

By integrating by parts and using the equation of motion,
the interaction Hamiltonian can be rewritten without
picking up boundary terms~\cite{rp_integ_by_parts}.
Using $(3.7)$ from~\cite{rp_integ_by_parts},
with $f=-\varepsilon/(c_s^2H)$,
we obtain the following form:
\begin{align}
    \Hint(t)=\ \int d^3x\bigg\{& -\frac{a^3\varepsilon}{H c_s^4}\left(-c_s^{2}-2c_s^{2}\frac{\lambda}{\Sigma}\right)\dot{\zeta}^3
		+ \frac{a^3\varepsilon}{c_s^{4}}\left(-3c_s^2\right) \zeta\dot{\zeta}^2\nonumber\\
		&- \frac{a\varepsilon}{c_s^{2}}\left(-c_s^2\right) \zeta(\partial\zeta)^2
        - \frac{a\varepsilon}{Hc_s^2}\dot{\zeta}(\partial\zeta)^2\nonumber\\
        &- \frac{a^3\varepsilon^2}{2c_s^{4}}(\varepsilon-4)\dot{\zeta}\partial\zeta\partial(\partial^{-2}\dot{\zeta})
        - \frac{a^3\varepsilon^3}{4c_s^4}\partial^2\zeta(\partial(\partial^{-2}\dot{\zeta}))^2\bigg\}.\label{interaction_eql}
\end{align}
To leading order, this formulation is made up of terms that
give equilateral shapes when the slow-roll parameters are roughly constant.
It was pointed out in~\cite{Funakoshi} that
using~\eqref{interaction_loc} in a scenario that results in an equilateral shape
would require sensitive cancellations in the squeezed limit.
Likewise, using~\eqref{interaction_eql} for a local scenario
would require sensitive cancellations in the equilateral limit.

As mentioned in~\cite{Funakoshi}, the spatial derivatives
can be manipulated into simple prefactors of $k_i$ using the
triangle condition ($\mathbf{k_1}+\mathbf{k_2}+\mathbf{k_3}=0$),
and so preserve the separability of the result.
To absorb these prefactors in our calculation, we precompute
$k^{p}q_a(k)$ as a linear combination of the $q_a(k)$ for the
relevant values of $p$,
from which $V^{(i)}_{P\tilde{P}}$ defined in~\eqref{V_definition} is built.
For certain sets of basis functions this matrix can be calculated analytically,
but it is simpler and more robust to numerically calculate the relevant integral directly.
The processing cost this incurs is small, and must only be
paid once per basis. We note especially that this means the matrix can
be stored and efficiently used in many scenarios.
To summarise,
we calculate the bispectrum contribution from each vertex in $\Hint$ separately:
we assemble the integrands, integrate them with respect to time,
include the prefactors coming from the spatial derivatives,
then sum the resulting sets of basis coefficients.
Of course, these methods are not restricted to this example of $\Hint$.

\subsection{Numerical considerations}\label{sec:k_dep}
The previous two sections detailed the methods required to calculate
the coefficients $\alpha_n$ in~\eqref{goal}, obtaining an
explicitly separable expression for the shape function.
In this set-up,
there are two kinds of integrals we must compute: integrals over time of
the form~\eqref{inin_kindep}, and integrals over $k$ of the form~\eqref{mode1Dcoeffs_integral}.
The first is done once per coefficient for each vertex, the second is a decomposition
done once for $\zeta_k$ and $\zeta_k'$ each, every timestep.
In this section we detail how to numerically evaluate these
integrals accurately and efficiently.

Since calculating each point in the time integrand requires a
decomposition~\eqref{mode1Dcoeffs_integral}, which is highly oscillatory
at early times, it is worthwhile to consider how to perform the
time integral efficiently. From the form of the Bunch-Davies mode functions,
we expect the dominant frequency (in $\tau_s$) to be $3\kmax$.
Assuming we are earlier than any features that might change this,
we can use this knowledge to sample the integrand at a far lower rate,
building the oscillation into our quadrature weights.
A second important consideration comes from how early we sample the integrand.
We can of course only obtain a point in the time integrand after our mode
functions have burned in from their set initial conditions to their true
attractor trajectory. This means that sampling earlier in the time integrand
requires us to set the initial conditions for the mode functions deeper
in the horizon, a regime in which they are expensive to evolve.

The integrals of the form~\eqref{inin_kindep} that we must calculate have $\tau=-\infty$ as their lower limit.
The highly oscillatory nature of the mode functions in these early times ($\lvert k\tau_s\rvert\gg1$) suppresses the
coefficients of our basis expansion by a factor of $1/\tau_s$.
As noted in~\cite{Funakoshi}, this means that we do not need to
explicitly use the $i\varepsilon$ prescription to force the integrals to converge.
In the case of using the Legendre polynomials as our
basis, we can see this more precisely by considering
the plane wave expansion (e.g.~\cite{finite_ft_legendre}):
\begin{align}\label{exp_expansion}
    e^{-i\bar{k}(\kmax-\kmin)\tau/2} = \sum_{n=0}^{\infty}(2n+1)i^n P_n(\bar{k})j_n(-(\kmax-\kmin)\tau/2)
\end{align}
for $\bar{k}$ in $[-1,1]$. When $(\kmax-\kmin)\tau/2$ is large, the spherical Bessel functions
oscillate with an amplitude $\propto\frac{1}{\tau}$. Thus,
the initial conditions~\eqref{bd_ic} expanded in Legendre polynomials (and similar)
give us suppression of $1/\tau^3$ in~\eqref{inin_kindep}.

While our method has extra suppression compared to configuration-by-configuration methods
(and thus does not need the $i\varepsilon$ prescription to converge)
it still converges rather slowly, as we push the lower limit to earlier times.
This expensive sampling can be wasteful of resources,
especially in a feature scenario where we know this
region will not contribute to the final result.
Care is required however, as starting the integration in the wrong way can easily lead to
errors which can completely swamp the result, since higher order modes
are more sensitive to early times.
The authors of~\cite{chen_easther_lim_1} used an artificial damping term to smoothly
``turn on'' their integrand.
The point at which this is done can then be pushed earlier to check for convergence.
However they found that the details of the damping needed to
be carefully set to avoid underestimating the result.
In~\cite{chen_easther_lim_2} they replaced this method by a ``boundary regulator'';
they split the integral into early and late parts and used integration by parts to
efficiently evaluate the early time contribution.
As our integrand already has extra suppression compared to the configuration-by-configuration
integrands considered in~\cite{chen_easther_lim_1,chen_easther_lim_2},
we can safely use the simpler first method.

We understand this situation by
taking advantage of asymptotic behaviour of highly oscillatory
integrals (for a review see~\cite{iserles_2005}).
Since the leading order term depends on
the value of the non-oscillatory part only at the endpoints,
and the next-to-leading order correction depends on the
derivative only at the endpoints, we can approximate the integral
$\int_{-\infty}^T f(\tau_s)e^{iw\tau_s}d\tau_s$
by replacing the non-oscillatory part $f(\tau_s)$ with a function with
matching value and derivative at $\tau_s=T$,
but which converges far faster.
We use
\begin{align}\label{smooth_cutoff}
f(\tau_s)e^{-\beta^2(\tau_s-T)^2},
\end{align}
for $\tau_s<T$.
In this way, for sufficiently large $T$,
we obtain the accuracy of the
first two terms of the asymptotic expansion
($O(\beta^2/w^2)$, $w=3\kmax$) without needing to explicitly
calculate the derivative at $T$, or needing any phase information (as
one would need to accurately impose a sharp cut on the integrand).

\begin{figure}[!pth]
\centering
\includegraphics[width=\columnwidth]{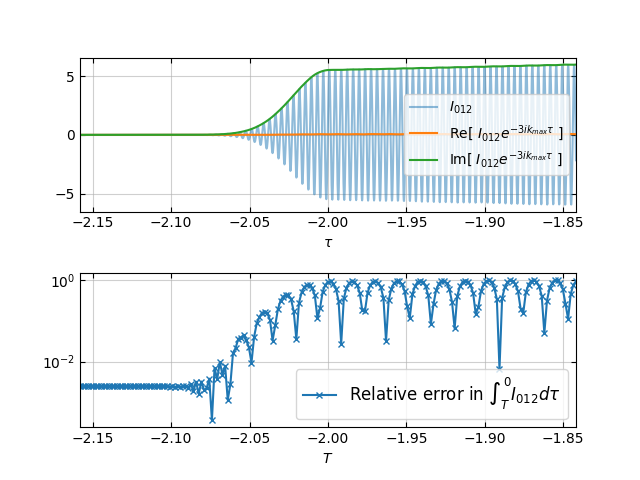}
\caption{
    A toy example demonstrating the considerations involved in performing the
    time integrals~\eqref{inin_kindep}.
    By carefully starting the time integrations,
    using the form~\eqref{smooth_cutoff}, we can avoid
    errors that would otherwise swamp our result.
    The coefficient being calculated is the
    $\alpha_{012}$ coefficient of the $\Lbasic$ expansion of~\eqref{example_eft2}.
}\label{time_integrands}
\end{figure}

We use a damping of the form $e^{-\beta^2(\tau_s-T)^2}$ for $\tau_s<T$ to smoothly
set the integrand to zero before a certain initial time, $T$.
As long as $T$ is sufficiently early and $\beta$ is not too large
their precise values have no significant effect on the final result.
For definiteness, we take $\beta/w=1\times10^{-4}$, small enough that
the integrand has many oscillations while it is ``turning on'',
so matches the contribution of an infinite limit to high accuracy.
We demonstrate this in figure~\ref{time_integrands}
for a toy $\Hint=(-1/\tau)\dot{\zeta}^3$,
as in~\eqref{example_eft2}.

To obtain a $k$-sample we must evolve a Fourier mode from Bunch-Davies initial conditions deep in the horizon
until it becomes constant after horizon crossing.
We denote by $N_k$ the number of Fourier modes we evolve.
Different choices of distributing the k-samples are possible; for example, one could distribute them
with an even spacing, log-spacing or cluster them more densely near $\kmin$ and $\kmax$.
The $k$-integrals themselves can be computed quite efficiently since
at every timestep the integral is over the same sample points.
One can therefore calculate and store the values of the basis functions at each of these points,
along with the integration weights which will depend on the distribution of $k$-samples.
The actual integration at each timestep, the calculation of the 
coefficients in~\eqref{mode1Dcoeffs_integral}, then becomes nothing more
than a dot product of a time-independent array with the
numerically evolved mode functions, for each order up to $\Pmax$.
We have found the best convergence results from distributing the k-samples
according to the prescription of Gauss-Legendre quadrature.

To calculate the basis expansion of the bispectrum using the in-in formalism
we must first calculate the basis expansion of the mode
functions at each timestep~\eqref{mode1Dcoeffs_integral}.
At early times the mode functions are highly oscillatory, taking the form $z_k e^{-ik\tau_s}$
for some much smoother $z_k$.
Directly decomposing this would require evolving more $\zeta_k$ samples
than is practical.
We want an expansion of the form
\begin{align}\label{mode_expansion}
	z_k e^{-ik\tau_s} = \sum_{n=0}^{\infty}\alpha_n q_n(k).
\end{align}
We can obtain this by using standard oscillatory
quadrature, if the $\tau_s$ dependence of the weights does not add too much overhead.
We can also use an expansion of $e^{-ik\tau_s}$
with a known explicit time dependence, for example
the expansion~\eqref{exp_expansion}.

To use this second method,
the first (smooth) factor $z_k$ can be expanded in whatever basis we are working in, $q_n(k)$,
and the second factor (highly oscillatory in $k$) is expanded
in some convenient basis $\tilde{q}_n(k)$
(e.g.\ $\Fbasic$, or $\Lbasic$ using the analytic form~\eqref{exp_expansion}).
Then by precomputing
$q_a(k)\tilde{q}_b(k)$
as a linear combination of the set of basis functions $q_c(k)$
all we need calculate at each timestep is the coefficients
of the smoother $z_k(\tau_s)$, which we then convert to the coefficients
of $F^{(i)}(\tau, k)$. In this way we can retain flexibility in our
bispectrum basis, as well as efficiency and precision in the calculation.
In the case of using $\Lbasic$ for the $\tilde{q}_n(k)$,
assuming the expansion in~\eqref{shapemodeexp} converges,
we need only compute the expansion for $e^{-ik\tau_s}$ to enough
terms that the first $\Pmax$ of the
coefficients in the expansion~\eqref{mode1Dcoeffs_sum} of the $F^{(i)}(\tau, k)$ converge,
not until the actual sum~\eqref{mode1Dcoeffs_sum} converges, since for high enough orders the integrals
in~\eqref{inin_kindep} will integrate to zero.

Clearly, once $\tau_s$ becomes small enough these considerations will no longer be necessary
and we can simply decompose the mode function directly.
We do this around the horizon crossing of the geometric mean of $\kmin$ and $\kmax$.
If there is an extreme feature which causes a large deviation from the usual slow-roll form
this switch will need to be made sooner. 
Also, this method would need to be adapted for non-Bunch-Davies initial conditions.
Since anything related to the basis but independent of the scenario can be
precomputed, certain parts of this calculation do not hurt the efficiency of this
method in the context of, for example, a parameter scan.
Using the methods outlined above,~\eqref{inin_kindep} and~\eqref{mode1Dcoeffs_integral}
can be computed precisely and efficiently in a mostly basis-agnostic context
allowing us to ({\romannumeral 1}) preserve the intrinsic separability of the tree-level
in-in formalism and ({\romannumeral 2}) do so in a way that allows easy exploration of possible
sets of basis functions, to find a set that converges quickly enough to
be useful in comparison with observation.

\section{Validation}\label{sec:validation}
\subsection{Validation methods}\label{sec:validation_methods}
In this section we validate our implementation of our methods
on different types of non-Gaussianity, sourced in different ways.
While our actual results take the form of a set of mode expansion coefficients $\alpha_n$,
to make contact with previous results in the literature
all of our validation tests take place on the tetrapyd,
the set of physical bispectrum configurations.

We test that our results have converged using~\eqref{relative_difference},
between $\Pmax=45$ and $\Pmax=15$ for the featureless cases,
and between $\Pmax=65$ and $\Pmax=35$ for the cases with features.
We will refer to this as our convergence test.
To verify that our results have converged to the correct shape,
we perform full tetrapyd checks against known analytic results
(where those are available, and in their regimes of validity)
using~\eqref{relative_difference},
and point tests against the PyTransport code for the scenarios
with canonical kinetic terms.
Since all our scenarios are single-field, the most general
test we have is the single-field consistency relation,
which states that for small $k_L/k_S$, the shape function $S(k_S,k_S,k_L)$
must obey~\eqref{eq:sqz_consistency}.
The consistency condition should hold most precisely
at the configurations with smallest $k_L/k_S$,
the most squeezed being the three corners, $(\kmax,\kmax,\kmin)$ and permutations.
We want our test to be on an extended region of the tetrapyd however,
so we choose the line
\begin{align}\label{sqz_line}
    \frac{k_L}{k_S}&=\frac{2\kmin}{\kmax},
\end{align}
which connects $(\kmax,\kmax,2\kmin)$ to $(\kmax/2,\kmax/2,\kmin)$.
We will take $\frac{\kmin}{\kmax}=\frac{1}{550}$, so this is still sufficiently squeezed to be a stringent test.

First, we investigate convergence on simple featureless models,
both local-type~\eqref{eq:quadratic_potential}
and equilateral-type~\eqref{eq:dbi_warp}.
We find that in our chosen basis $\Lnsboth$ our results
converge quickly and robustly as we increase the number of modes,
where we quantify the convergence using~\eqref{relative_difference}.
We compare the converged results against analytic
templates~\eqref{malda_shape} and~\eqref{dbi_shape},
using the full shape information~\eqref{relative_difference},
finding them to match to high accuracy.
Secondly we validate our methods on an example of non-Gaussianity
from a feature: linear oscillations from a sharp step in the
potential~\eqref{eq:kink_potential}. The result converges robustly
across the parameter range we explore. Throughout that range,
we test the converged result using the squeezed limit consistency
condition, and perform point tests against PyTransport,
finding excellent agreement.
For small step size we can further validate against the analytic template
of~\cite{adshead}, using the full shape information, finding agreement
to the expected level given the finite width of the step.
The final type of non-Gaussianity we use for validation on is the resonance
type, logarithmic oscillations generated deep in the horizon~\eqref{eq:resonant_potential}.
We test the converged result against the PyTransport
code, by performing point tests on a slice.
We also present a resonant DBI scenario, with out-of-phase oscillations
in the flattened limit, as pointed out in~\cite{chen_folded_resonant},
resulting from non-Bunch-Davies behaviour of the mode functions.
We also test both resonant scenarios using the squeezed limit consistency condition.

\begin{figure}[!pth]
\centering
    \subfloat{\includegraphics[width=.45\columnwidth]{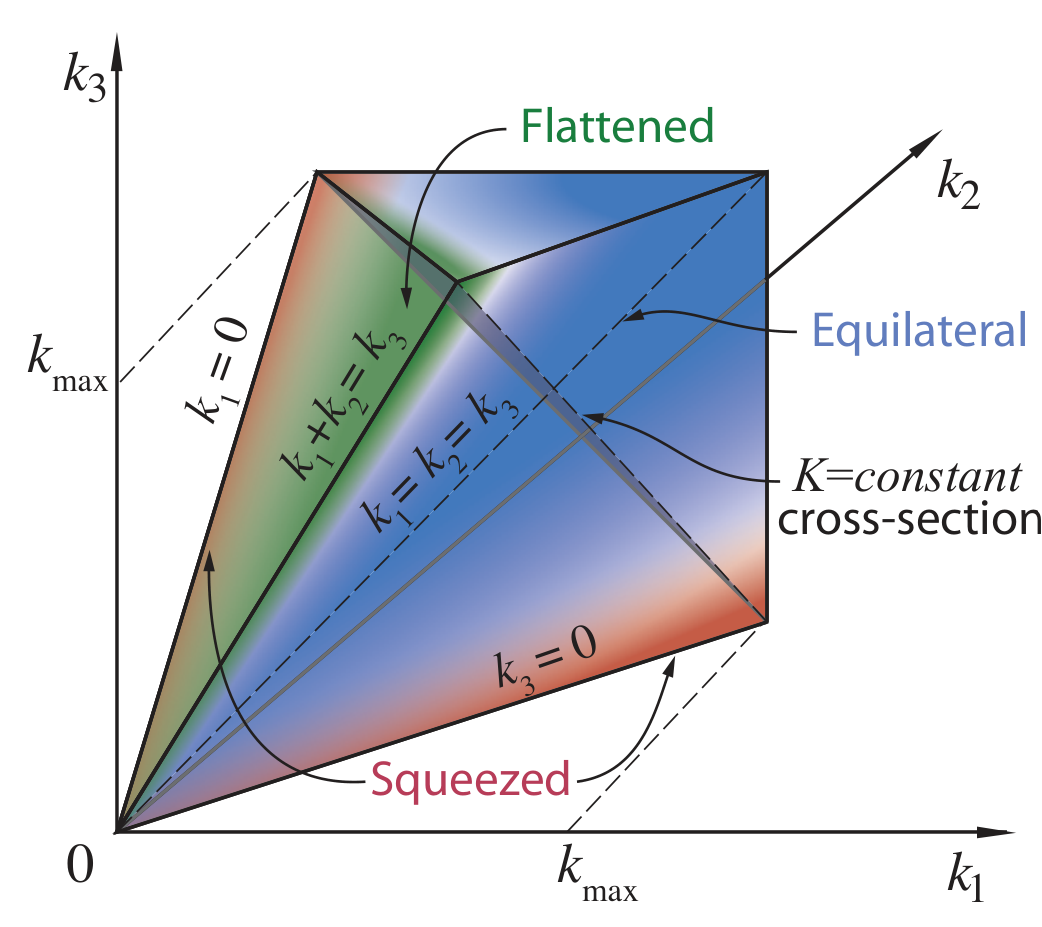}}
    \subfloat{\includegraphics[width=.55\columnwidth]{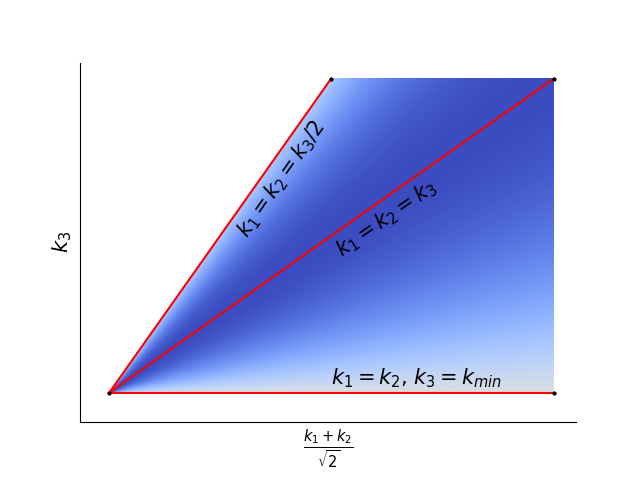}}
\caption{
    For ease of display, we will plot the two-dimensional $k_1=k_2$ slice of the tetrapyd
    for each of our validation examples, as shown schematically here on the right.
    Horizontal lines on this plot have
    constant $k_3$. The bottom edge is $k_3=\kmin$, the top
    edge is $k_3=\kmax$. The right edge is $k_1=k_2=\kmax$, the
    left edge is $k_1=k_2=k_3/2$, i.e.\ the limit imposed by the triangle condition.
    Plotted in red (in the right hand plot) from top-left to bottom-right,
    are the flattened, equilateral and squeezed limits. For comparison,
    half of the tetrapyd is shown in the three-dimensional plot on the left.
}\label{slice_explained}
\end{figure}

We display the phenomenology of our various validation examples
by plotting slices through the tetrapyd, as detailed in figure~\ref{slice_explained}.
Along with the phenomenology plots we plot the residual
(with respect to the totally converged result)
on the same slice, relative to the magnitude of the shape~\eqref{rep_val}.
We emphasise that while these plots display slices through the
tetrapyd, our actual result describes the shape function on the
entire three-dimensional volume of the tetrapyd, and
we measure our convergence over this whole space.

While one of the main advantages of this method is its direct link
to the CMB, in this section we only concern ourselves with validating the
code, not the observational viability of the scenarios considered.
We focus on accurately and efficiently calculating the primordial
tree-level comoving bispectrum, validating on models popular in the
literature.
\subsection{Quadratic slow-roll}
\begin{figure}[!pth]
\centering
    \subfloat{\includegraphics[width=0.99\columnwidth]{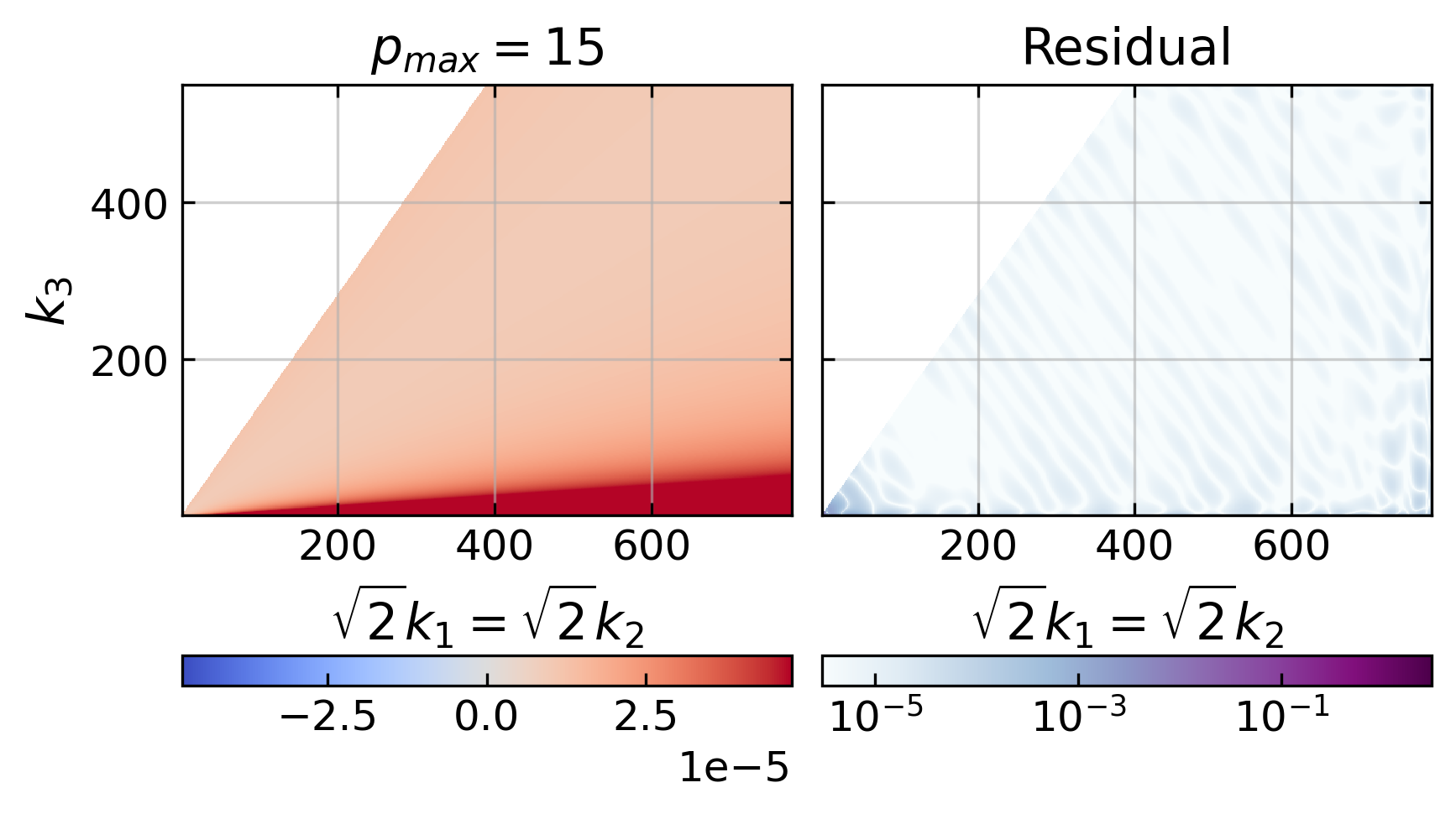}}\\[-2ex]
    \subfloat{\includegraphics[width=0.99\columnwidth]{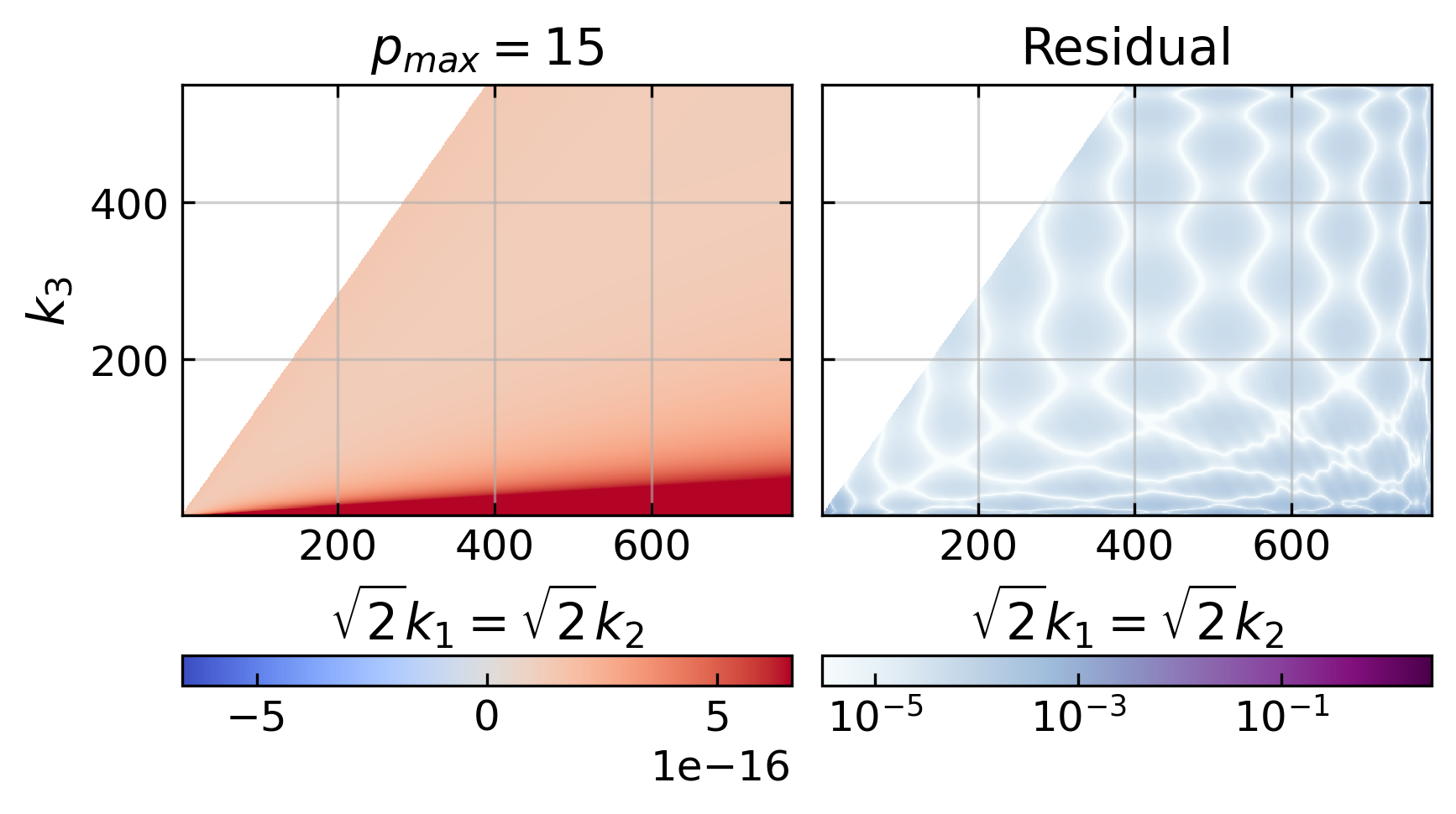}}
\caption{
    A canonical single-field model on a quadratic
    potential~\eqref{eq:kink_potential},
    slowly-rolling with $\varepsilon\approx2\times10^{-6}$
    in the top plot, and $\varepsilon\approx0.8\times10^{-2}$
    in the lower plot.
    This shape is dominated by its squeezed limit,
    and has a scale dependence determined by $\varepsilon$,
    very small in the top plot and ``realistic'' in the
    lower plot, relative to the $\textit{Planck}$ power spectrum.
    The first scenario converges well in the $\Linvk$ basis,
    with a relative difference of $2.7\times10^{-5}$
    %2.6764e-05
    between $\Pmax=45$ and $\Pmax=15$.
    The second scenario converges well in the $\Lnsboth$ basis
    (with $n_s^{*}-1 = -0.0325$),
    with a relative difference of $7.9\times10^{-5}$
    % 7.9047e-05
    between $\Pmax=45$ and $\Pmax=15$.
}\label{slice_plot_malda}
\end{figure}

The first model we will consider is slow-roll inflation
on a quadratic potential~\eqref{eq:quadratic_potential}.
We consider two scenarios, both with $m=6\times10^{-6}$.
The first is deep in slow-roll, which we achieve by choosing
$\phi_0=1000$; then, choosing $\phi'_0$ according to the
slow-roll approximation, we get
$\frac{1}{2}\phi'^2=\varepsilon\approx0.2\times10^{-5}$.
We can then choose the initial value for $H$ to
satisfy the Friedmann equation to sufficient precision.
The second scenario is chosen to have a value for $n_s^{*}-1$
consistent with the $\textit{Planck}$ result,
by choosing $\phi_0=16.5$, so that $\varepsilon\approx0.8\times10^{-2}$.
The shapes are shown in figure~\ref{slice_plot_malda}.

We choose the first scenario to have such a small value of $\varepsilon$
so that we can use Maldacena's shape~\eqref{malda_shape}
as a precision test.
Indeed, we find that it has a scaled relative difference~\eqref{relative_difference_scaled}
of $2.7\times10^{-5}$ with this shape,
contrasting a scaled relative difference
of $0.077$ with the local template~\eqref{local_shape}.
This confirms that our methods and our implementation in code can accurately
pick up this basic type of featureless non-Gaussianity.

For the second scenario, we cannot validate on
Maldacena's shape~\eqref{malda_shape} or the
local template~\eqref{local_shape},
as for $\epsilon\approx0.8\times10^{-2}$ we only expect
these templates to match the true result to percent level accuracy.
Indeed, we find that our result has a correlation of $0.998$
with both~\eqref{malda_shape} and~\eqref{local_shape},
corresponding (in the sense of~\eqref{relative_difference_scaled})
to a relative difference of $6\%$,
as expected.
Instead, we validate this model using
the squeezed limit test described above,
verifying our result to $0.05\%$.

This is a validation of the convergence of our basis,
reaffirming the template decomposition results of figure~\ref{fig:recon_malda_dbi}
in the setting of the in-in formalism.
It is also a stringent validation of our methods of including the higher-order
coefficients, as insufficient care taken in the early-time sections of
integrals~\eqref{inin_kindep}, or in including the spatial derivatives
from $\Hint$, could have easily swamped the $\Pmax=45$ result.

\subsection{DBI inflation}
\begin{figure}[!pth]
\centering
    \subfloat{\includegraphics[width=0.99\columnwidth]{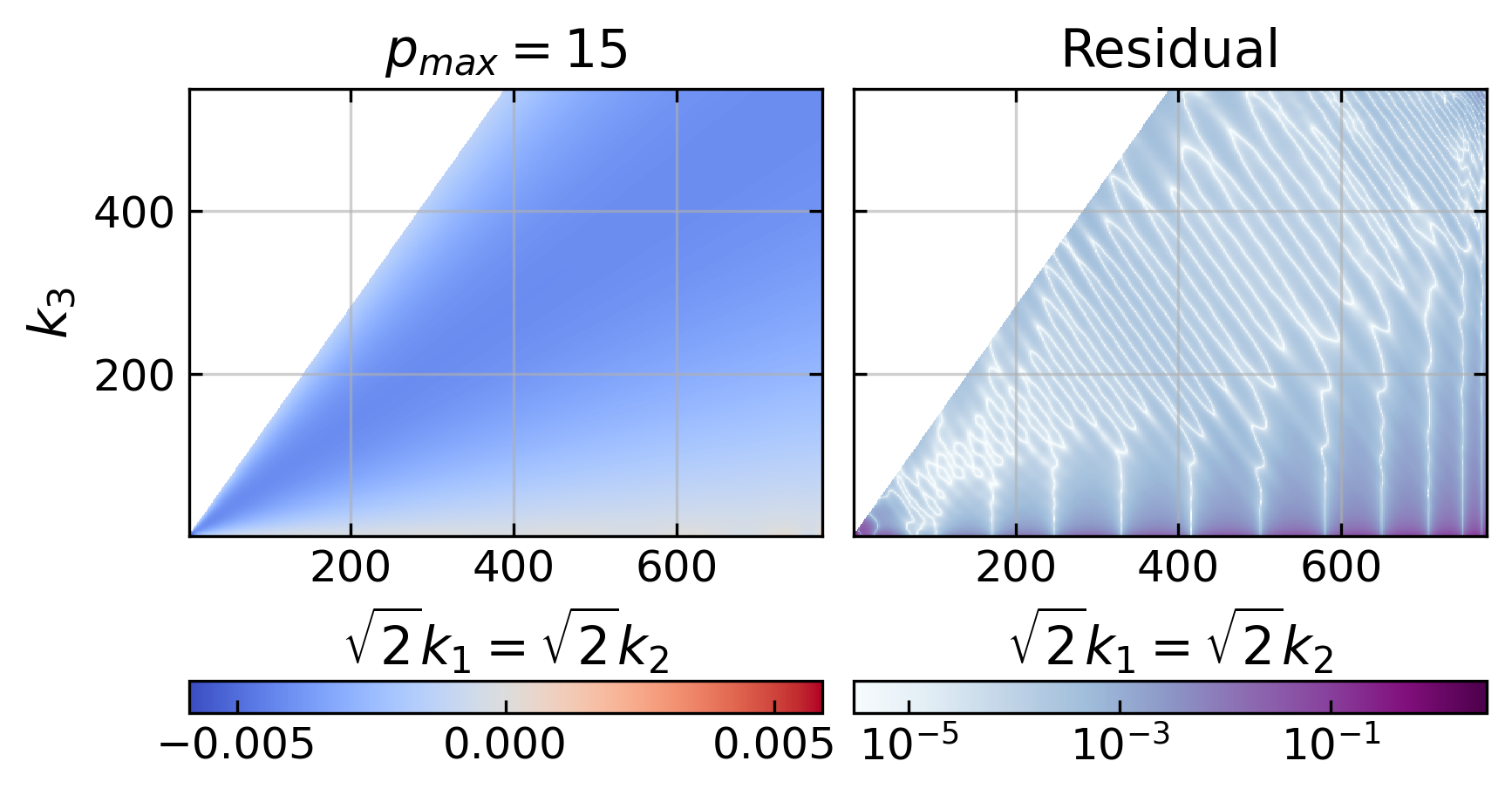}}\\[-2ex]
    \subfloat{\includegraphics[width=0.99\columnwidth]{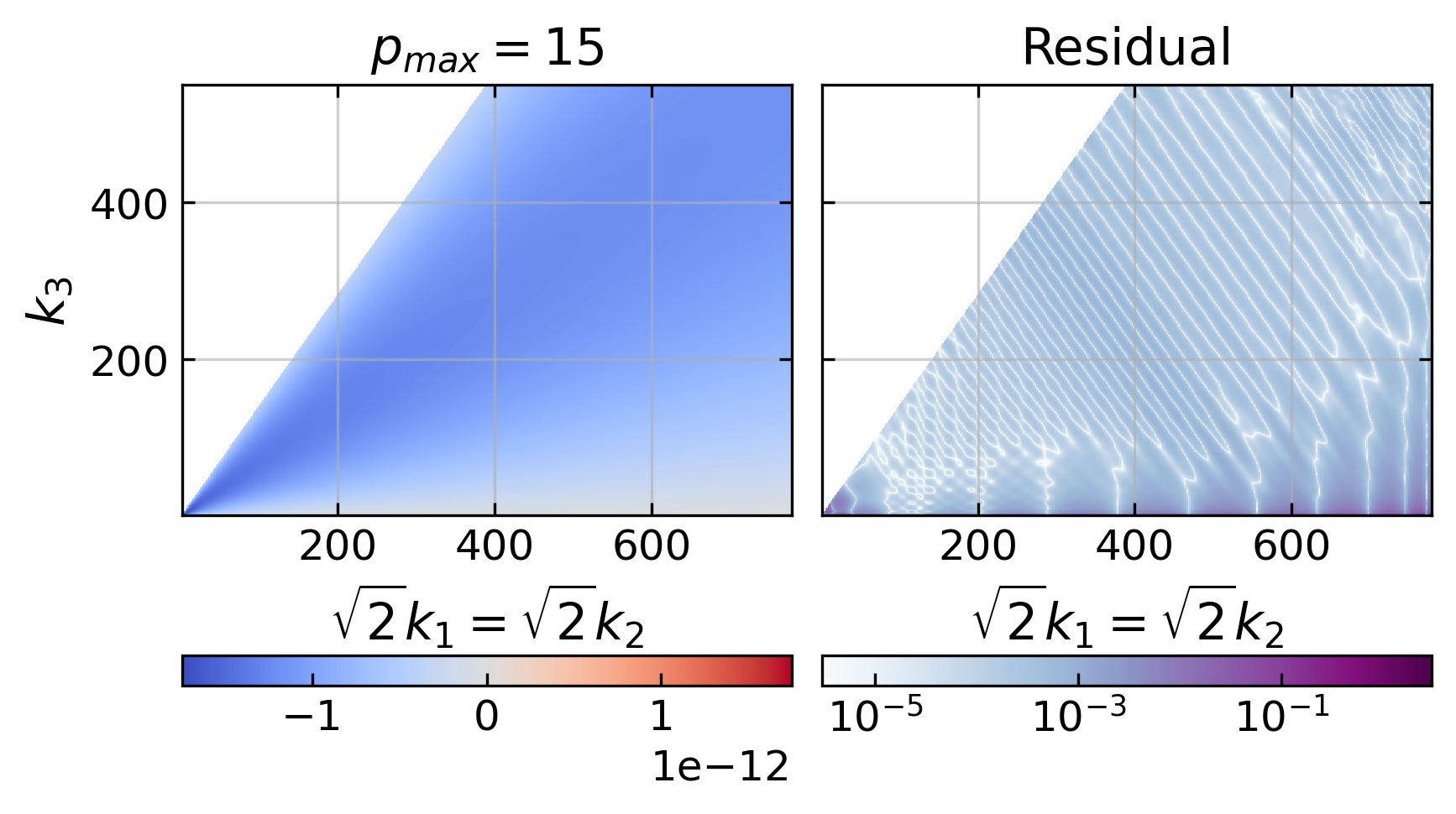}}
\caption{
    The upper plot shows the shape function for a DBI model
    deep in slow-roll. We set
    $\lambda_{DBI}$ in~\eqref{eq:dbi_warp} to $1.9\times10^{18}$,
    obtaining a scenario with $\varepsilon\approx1.9\times10^{-6}$ and
    $c_s=2.3\times10^{-3}$.
    This shape is dominated by its equilateral configurations,
    and has only a slight scale dependence.
    It converges well in the $\Linvk$ basis,
    with a relative difference of $2.1\times10^{-3}$
    % 2.1203e-03
    between $\Pmax=45$ and $\Pmax=15$.
    The lower plot shows a DBI model that saturates the {\it{Planck}}
    limit on $c_s$. We set
    $\lambda_{DBI}$ in~\eqref{eq:dbi_warp} to $1.9\times10^{15}$,
    obtaining a scenario with $\varepsilon\approx8.0\times10^{-5}$ and
    $c_s=8.0\times10^{-2}$.
    This shape is also dominated by its equilateral configurations,
    but has a scale dependence consistent with the measured
    power spectrum.
    It converges well in the $\Lnsboth$ basis
    (with $n_s^{*}-1 = -0.0325$),
    with a relative difference of $1.1\times10^{-3}$
    % 1.1461e-03
    between $\Pmax=45$ and $\Pmax=15$.
}\label{slice_plot_dbi}
\end{figure}

Next, we show results for a similar pair of scenarios for DBI inflation.
We choose $V_{0}={5.2\times10^{-12}}$~% and $\beta_{IR}=0.29$
with $m=\sqrt{0.29V_0/3}$
in~\eqref{eq:dbi_action} and~\eqref{eq:dbi_warp}.
We choose $\phi_0=0.41$, and then the starting condition
for $H$ according to the slow-roll approximation,
allowing us to choose $\phi'_0$ such that the Friedmann
equation is satisfied to sufficient precision.
The first scenario is deep in slow-roll, with $\lambda_{DBI}=1.9\times10^{18}$, while
the second scenario saturates the $\textit{Planck}$
limit on $c_s$, with $\lambda_{DBI}=1.9\times10^{15}$.
The resulting shapes are shown in figure~\ref{slice_plot_dbi}.

The scenario deep in slow-roll has a error of $0.082\%$
%0.0008243088
relative to the DBI shape~\eqref{dbi_shape},
and $13\%$
%0.1313763404
relative to the equilateral template~\eqref{equil_shape}.
The second scenario has a relative error of $2.9\%$
%0.0293218555
with the scale-invariant DBI shape, and $14\%$ with the equilateral template.
%0.1391026565
Including some scale dependence in the template,
using~\eqref{dbi_ns_shape}, we get a relative error of $0.27\%$.
%0.0026744179
On the line defined by~\eqref{sqz_line},
both scenarios have a sub-percent difference from the
consistency condition, with respect to the equilateral configurations,
which decreases when configurations with a larger $k_S/k_L$ are considered.

Including the minimal information of an individual, approximately
representative value of $n_s^{*}-1$
in $\Lnsboth$ allows us to converge to these smooth shapes quickly
and robustly, overcoming the tetrapyd-vs-cube difficulties described
in~\ref{sec:choice_of_basis}. Our accurate match to these shapes validates our
implementation in code, and the ability of the method
(and our basis in particular) to capture very different types of
bispectrum shapes, local and equilateral.

\subsection{Step features}
\begin{figure}[!pth]
\centering
    \subfloat{\includegraphics[width=0.99\columnwidth]{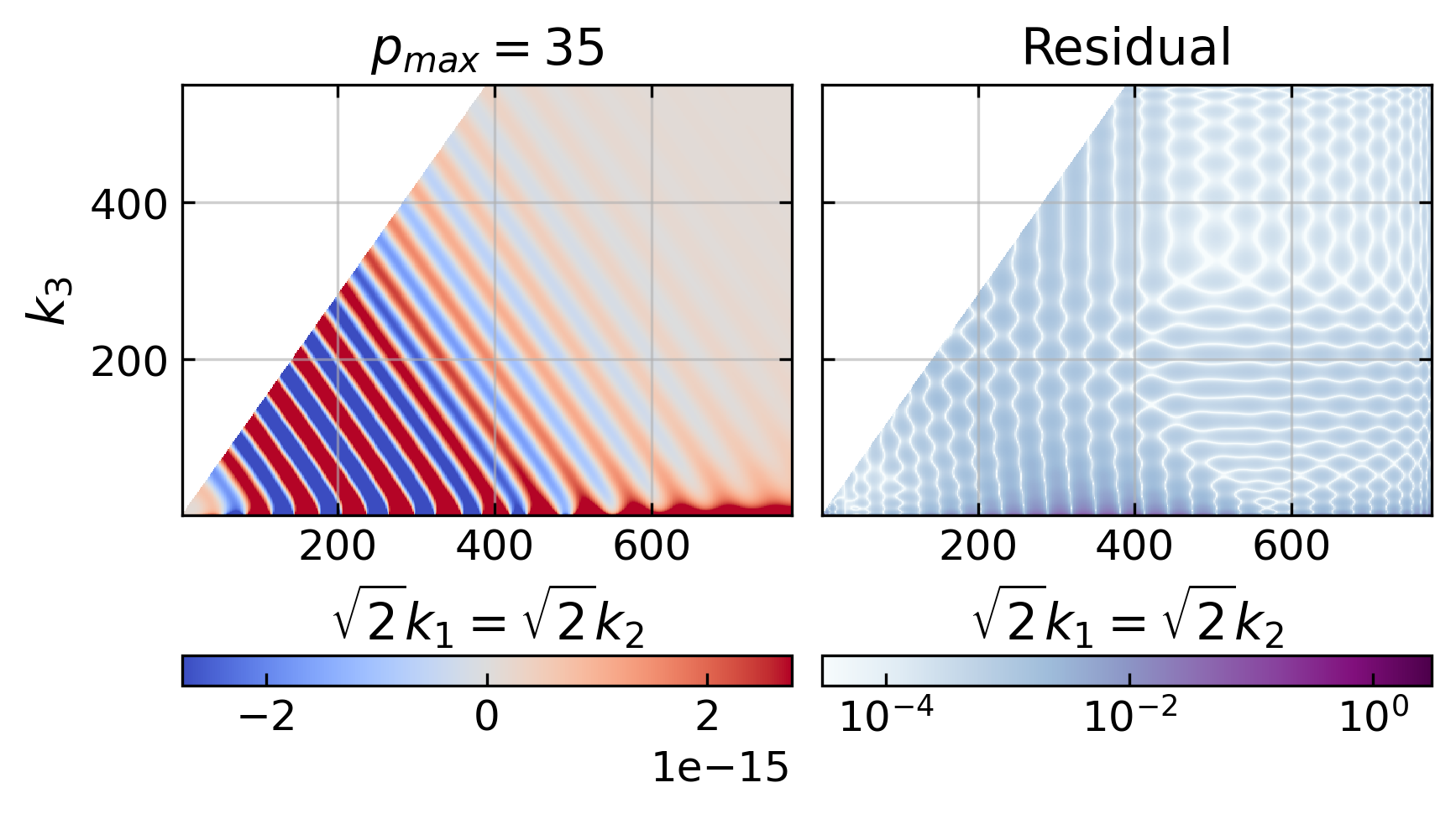}}\\[-2ex]
    \subfloat{\includegraphics[width=0.99\columnwidth]{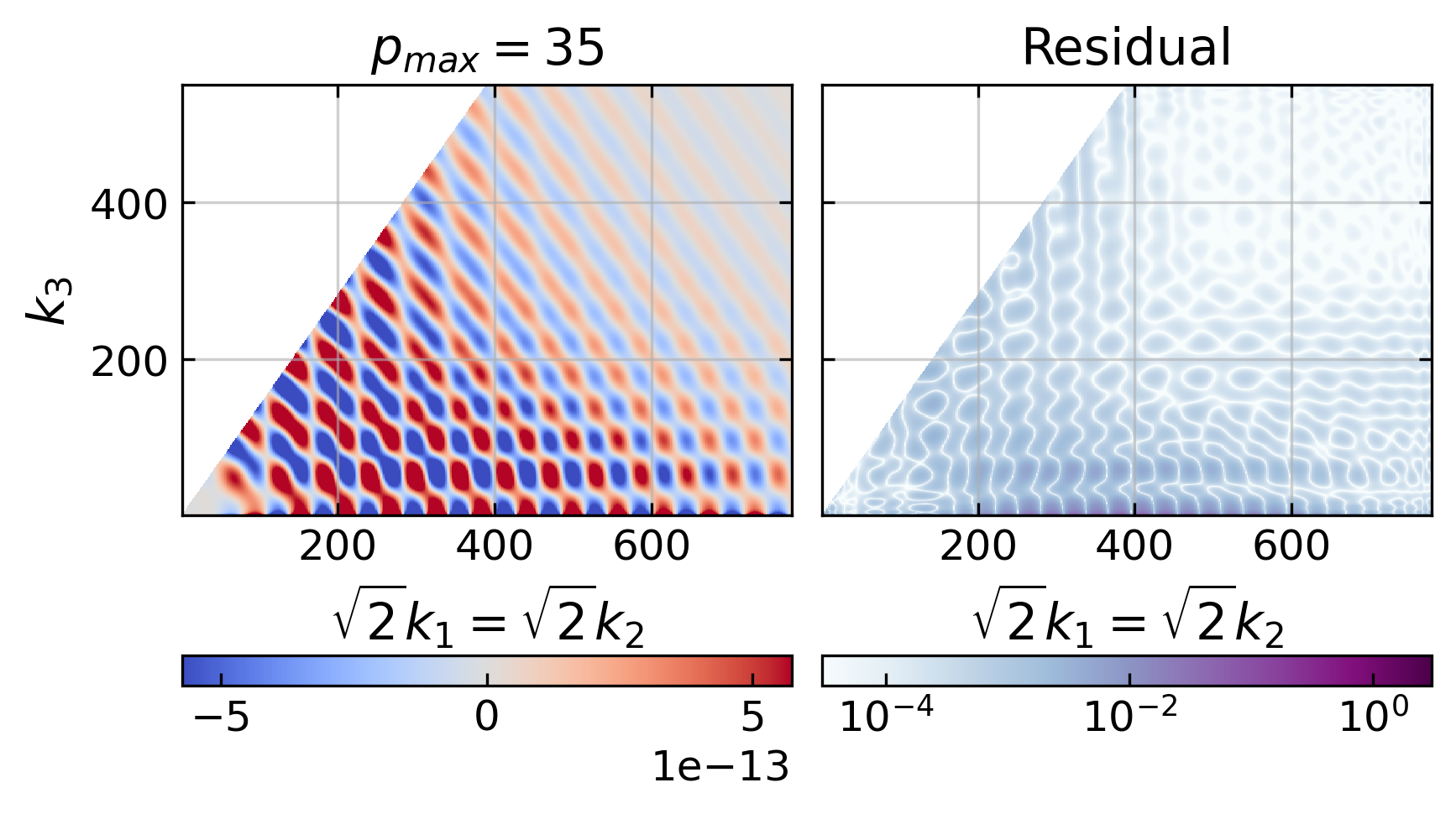}}
\caption{
    The tree-level shape function of a feature
    model~\eqref{eq:kink_potential}, shown for step sizes of
    $c=5\times10^{-5}$ (upper plot)
    and $c=5\times10^{-3}$ (lower plot).
    The corresponding expansion parameter values of~\cite{adshead},
    ${\mathcal{C}=6c/(\varepsilon+3c)}$, are $0.035$
    and $1.3$.
    For the smaller step size, the oscillations are almost entirely functions of $K=k_1+k_2+k_3$,
    except for a phase difference in the squeezed limit.
    The dependence is more complicated for $\mathcal{C}=1.3$,
    however our result still converges well.
    In the $\Lnsboth$ basis, with $n_s^{*}-1 = -0.0325$,
    the results have a relative difference of $1.6\times10^{-3}$
    and $1.5\times10^{-3}$, respectively,
    % 1.6406e-03
    % 1.6418e-03
    % 1.4856e-03
    between $\Pmax=65$ and $\Pmax=35$.
}\label{slice_plot_tanh}
\end{figure}

\begin{figure*}
    %\centering
    \subfloat{\includegraphics[width=0.49\columnwidth]{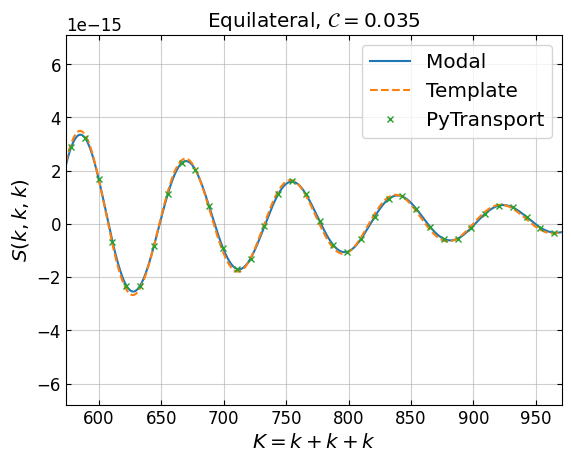}}
    \hfill
    \subfloat{\includegraphics[width=0.50\columnwidth]{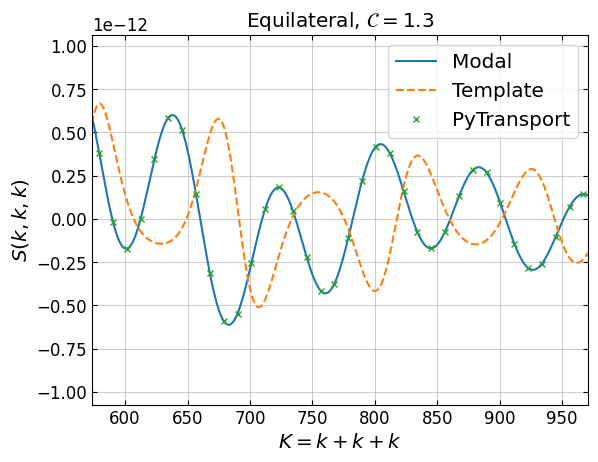}}
    \vskip\baselineskip
    \subfloat{\includegraphics[width=0.49\columnwidth]{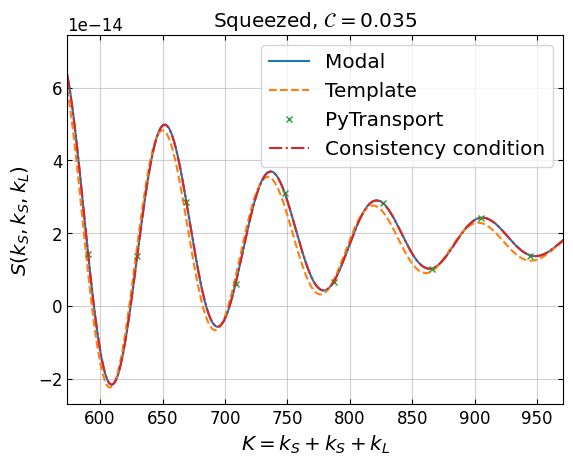}}
    \hfill
    \subfloat{\includegraphics[width=0.49\columnwidth]{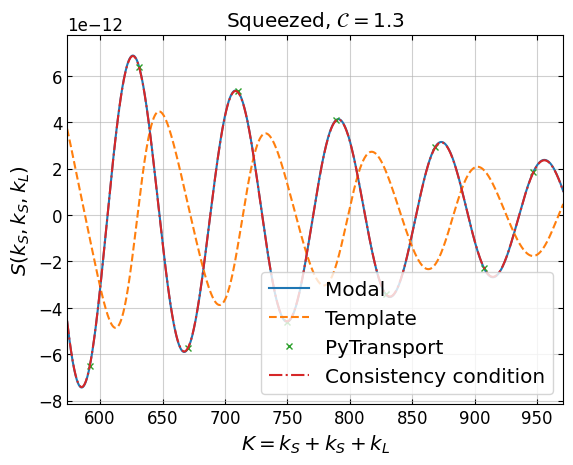}}
\caption{
	In the equilateral limit for the feature models (the top two figures) we validate our modal result
	against the PyTransport result. In the squeezed limit (the bottom two figures)
	we validate against PyTransport, and the consistency condition.
	In both limits, for both step sizes shown, we find excellent agreement.
	For the small step size (the two plots to the left), we additionally
	see a good match to the template of~\cite{adshead}. For the larger step size,
	the template amplitude is still accurate,
    but no longer captures the detailed shape information.
    This validates our code on non-Gaussianity generated by sharp
    features, and illustrates the general usefulness of our method.
	Our numerical results are accurate in a broader range than
	approximate templates, but are still smooth separable functions,
	unlike the results of previous numerical codes.
}\label{fig:tanh_sqz}
\end{figure*}

\begin{figure}[!pth]
\centering
\includegraphics[width=.75\columnwidth]{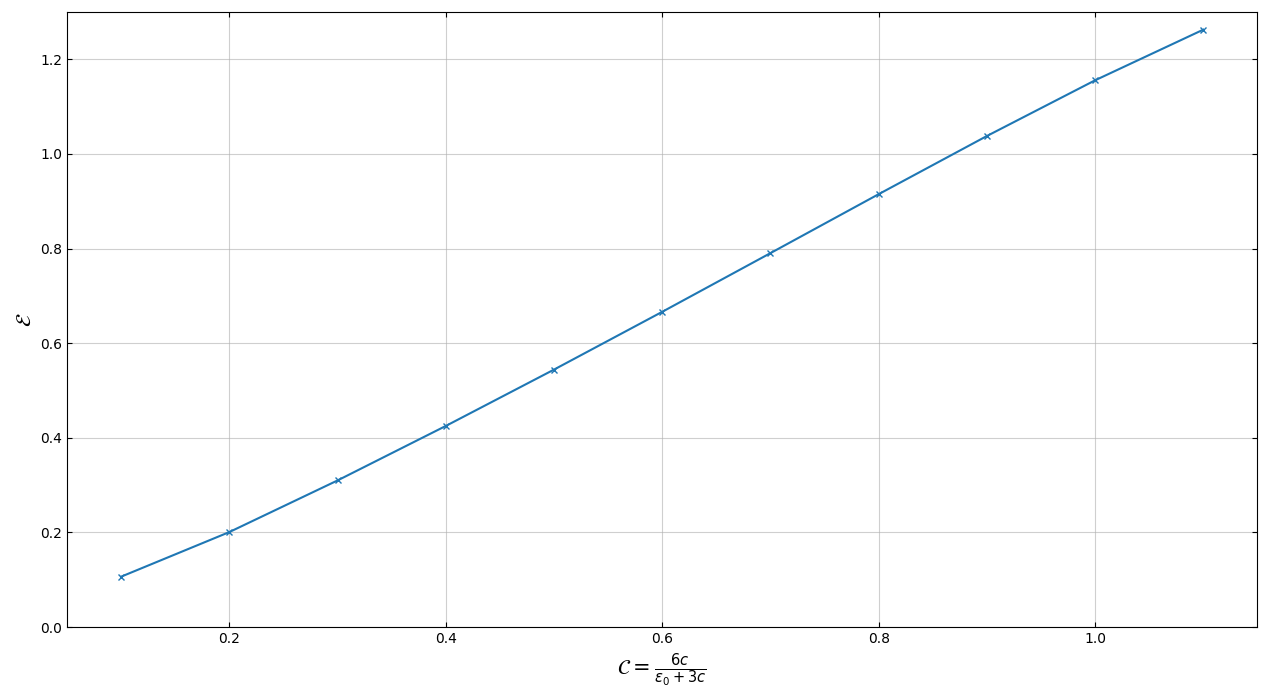}
\caption{
    We sample more shapes with step sizes between the two
    feature models shown in figure~\ref{slice_plot_tanh}.
    We plot the relative difference,
    integrated over the full tetrapyd
    in the sense of~\eqref{relative_difference},
    between the modal result and the analytic template of~\cite{adshead},
    as a function of the template parameter $\mathcal{C}=\frac{6c}{\varepsilon_0+3c}$
    (where $c$ is the step size and $\varepsilon_0$ is the value of the slow-roll parameter
    $\varepsilon$ at $\phi_{step}$ when $c=0$). We test our result by verifying
    the squeezed limit consistency
    condition to better than 1\% throughout (not shown). The number of oscillations in the
    $k$-range is determined by the conformal time at which the kink in~\eqref{eq:kink_potential}
    occurs, which is kept constant across this scan. The width of the feature was also kept constant.
}\label{fig:tanh_scan}
\end{figure}

Moving on from simple featureless bispectra, we present the results of
our validation tests on non-Gaussianity coming from a sharp feature in the potential.
We use the same parameters for the quadratic potential as in the
second scenario in fig~\ref{slice_plot_malda}.
In~\eqref{eq:kink_potential}
we fix $d=1\times10^{-2}$ and $\phi_{f}=15.55$
(as with the second canonical quadratic example, $\phi_0=16.5$).
Figure~\ref{slice_plot_tanh}
shows results for the shape function for two step sizes,
$c=5\times10^{-5}$ and $c=5\times10^{-3}$.
The resulting shape for small step sizes contains simple oscillations,
linear in $k_1+k_2+k_3$,
whose phase is almost constant across the tetrapyd.
When the step size is small, as expected,
our result matches the analytic result of~\cite{adshead},
presented there in equations (48), (54), (55).
We plot a comparison of the result of~\cite{adshead} and our result
in figure~\ref{fig:tanh_scan}.
For larger step size, we check the squeezed limit in figure~\ref{fig:tanh_sqz},
where we also show point tests against the PyTransport code.
Across this range of step sizes, for the resulting shapes we obtain
a full tetrapyd convergence test result
(between $\Pmax=65$ and $\Pmax=35$)
of between $0.17\%$ and $0.15\%$ and
we verify the squeezed limit test to better than $0.5\%$.

These examples show the utility of our methods in
calculating bispectra with non-trivial shape and scale dependence,
going beyond the simple examples of~\cite{Funakoshi}.
They validate the calculation of the high order coefficients,
and show that our code as implemented can handle sharp deviations from slow-roll,
generating non-Gaussianity around horizon crossing.


\subsection{Resonance features}
\begin{figure}[!pth]
\centering
\includegraphics[width=0.99\columnwidth]{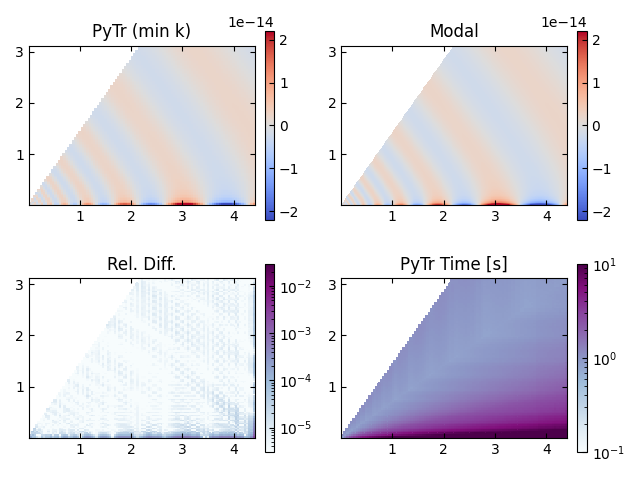}
\caption{
    Resonance on a quadratic potential~\eqref{eq:resonant_potential},
    testing our result using point tests against the
    PyTransport code.
    The logarithmic oscillations in the shape function are generated by periodic
    features deep in the horizon.
    The differences between our result and the PyTransport result
    are sufficiently small throughout that we can consider this
    a validation of our code on non-Gaussianity generated by periodic
    features deep in the horizon.
    In the $\Lnsboth$ basis, with $n_s^{*}-1 = -0.0325$,
    our result has a relative difference of $9.6\times10^{-3}$
    % 9.5587e-03
    between $\Pmax=65$ and $\Pmax=35$.
}\label{pytr_comparison_min}
\end{figure}

Now we further validate our code against two
resonance models. In contrast to the previous sharp kink,
this feature is extended, requiring precision at earlier times.
The first, shown in figure~\ref{pytr_comparison_min},
is a model with a canonical kinetic term, on a
quadratic potential with a superimposed
oscillation~\eqref{eq:resonant_potential}.
We take $bf=10^{-7}$, and $f=10^{-2}$.
The resulting bispectrum has oscillations logarithmic in $k_1+k_2+k_3$.
In figure~\ref{pytr_comparison_min} we see the excellent agreement
between our result and the PyTransport result, once initial conditions
in both codes are set early enough to achieve convergence.
This validates the code on non-Gaussianity generated deeper
in the horizon. Note the change of phase in the squeezed limit,
though this is expected to be unobservable.
We obtain a full tetrapyd convergence test result
(between $\Pmax=65$ and $\Pmax=35$)
of $0.93\%$, a squeezed limit test result of $1.1\%$
(along the line defined by~\eqref{sqz_line}),
and a relative difference of $3.0\%$ with respect to the
PyTransport result,
although this is only integrated over the two-dimensional slice presented in
figure~\ref{pytr_comparison_min}.

The time taken for the PyTransport code (per configuration) varies by a factor of around forty
between the equilateral limit and the squeezed limit,
as we show in figure~\ref{pytr_comparison_min}.
While the PyTransport code is extremely fast at calculating the shape function
for a single $k$-configuration,
to obtain this two-dimensional slice through the tetrapyd took around seven hours;
to obtain the shape function on the full three-dimensional tetrapyd would take much longer.
In contrast, our code took less than an hour on the same machine to calculate
the full shape function, not limited to the shown slice.
The overall speed increase is, therefore, a factor on the order of $10^2$ to $10^3$
for the full shape information, speaking only on the level of primordial
phenomenology, in addition to the advantage that our result is in a form
designed to be compared with observation.
We expect that our implementation can be optimised beyond this.

\begin{figure}[!pth]
\centering
\includegraphics[width=0.99\columnwidth]{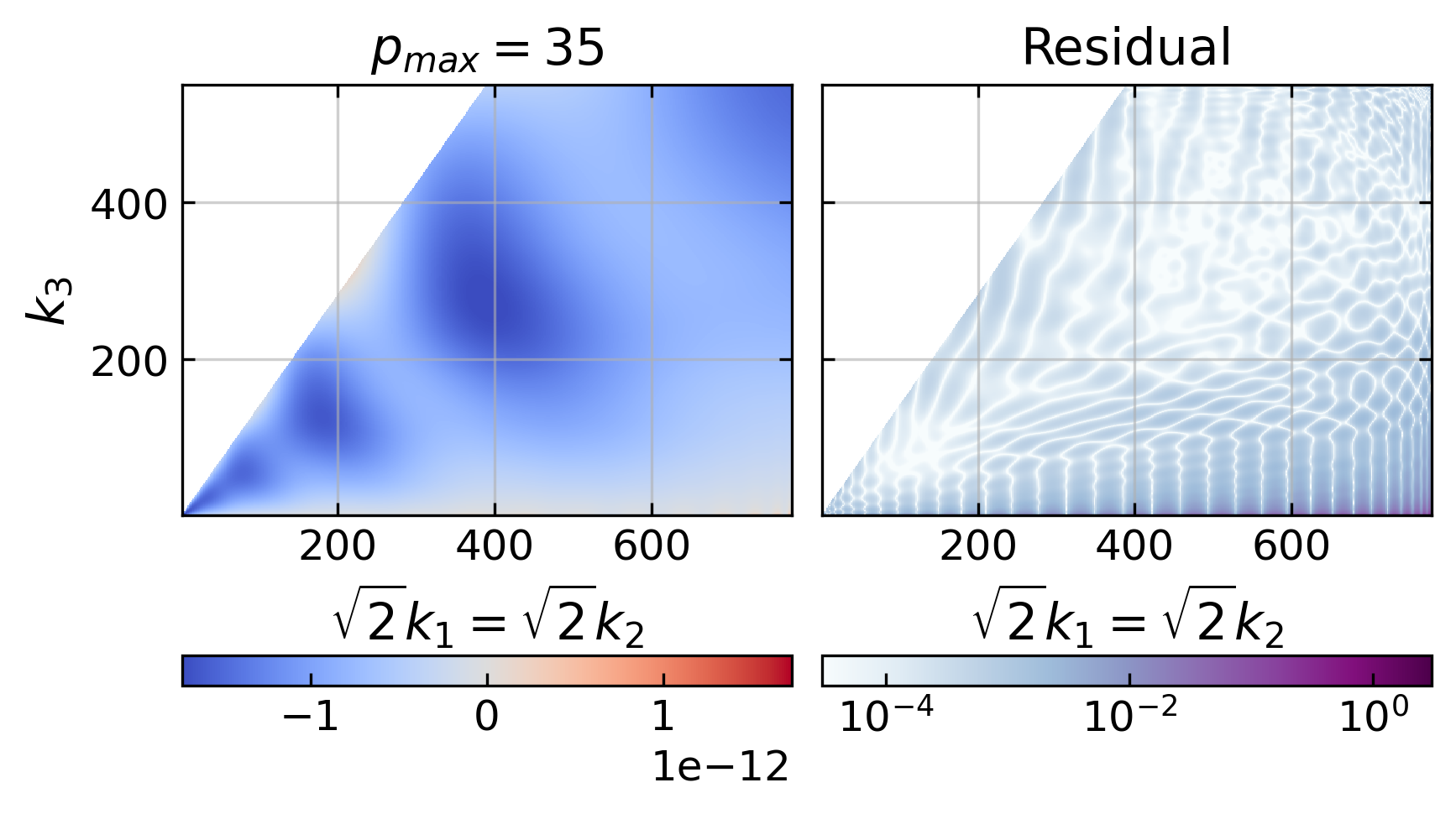}
\caption{
    Non-Gaussianity generated by periodic
    features in a DBI model, including a phase difference
    in the flattened limit as described in~\cite{chen_folded_resonant}.
    For the purposes of demonstrating the phenomenology,
    we have placed an envelope on the oscillations in the potential
    to aid convergence.
    In the $\Lnsboth$ basis, with $n_s^{*}-1 = -0.0325$,
    the result has a relative difference of $1.9\times10^{-3}$
    % 1.9482e-03
    between $\Pmax=65$ and $\Pmax=35$.
}\label{slice_plot_dbi_reso}
\end{figure}

The second scenario we consider here also has an
oscillation superimposed on its potential, but this time
is a non-canonical model, the DBI model.
The resulting bispectrum is shown in figure~\ref{slice_plot_dbi_reso}.
Note especially the out-of-phase oscillations in the flattened limit,
which are potentially observable.
For the purpose of displaying this phenomenology, we place a window
on the oscillation in the potential, smoothing out the resulting oscillations
in the shape at low $k_1+k_2+k_3$, to aid convergence.
This validates our code on non-Gaussianity generated by deviations from Bunch-Davies
behaviour~\cite{chen_folded_resonant,features_bartolo}.
We obtain a convergence test result (between $\Pmax=65$ and $\Pmax=35$)
of $0.15\%$, and a squeezed limit test result of $6.5\%$.

\section{Discussion}\label{sec:future}
In this work we have extended the modal methods of~\cite{FergShell_1,FergShell_2,FergShell_3}
to recast the calculation of the tree-level primordial bispectrum~\eqref{in-in}
into a form that explicitly preserves its separability.
We emphasise again that this work has two main advantages over previous
numerical methods. The more immediate is that by calculating the primordial
bispectrum in terms of an expansion in some basis, the full bispectrum can
be obtained much more efficiently than through repetitive integration
separately for each $k$-configuration. The second, more important, advantage
is the link to observations.
Unlike previous numerical and semi-analytic methods,
once the shape function is expressed in some basis
as in~\eqref{shapefn},
the integral~\eqref{eq:reduced_cmb} and other computationally intensive steps involved
in estimating a particular bispectrum in the CMB, can be precomputed. Since this
large cost is only paid once per basis, once a basis
which converges well for a broad range of models
has been found, an extremely broad exploration of primordial bispectra becomes immediately feasible in the CMB.
Making explicit the $k$-dependence in this way also opens the door to vast increases in efficiency in
connecting to other observables, by precomputation using the basis set, then performing a (relatively) cheap
scan over inflation parameters.

Our work here goes beyond that of~\cite{Funakoshi} in that our careful methodology
allows us to accurately and efficiently go to much higher orders,
in particular our methods of starting the time integrals~\eqref{inin_kindep}
and of including the spatial derivative terms in the calculation.
This allowed us to present this method for feature bispectra for the first time,
demonstrating the efficient exploration of much more general primordial bispectrum phenomenology.
We have also identified and addressed the effects
of the non-physical $k$-configurations on convergence within the three-dimensional tetrapyd.
We explored, for the first time, possible basis set choices in the context
of those effects.
We showed rapid convergence on a broad range of scenarios,
including cases with oscillatory features with non-trivial shape dependence,
using our augmented Legendre polynomial basis, $\Lnsboth$.

The immediate application of this work is the efficient exploration of
bispectrum phenomenology, as our methods can much more quickly
converge to the full shape information than previous numerical methods,
which relied on calculating the shape function point-by-point, for each $k$-configuration separately.
We have implemented these methods for single field scenarios
with a varying sound speed, scenarios which
have a rich feature phenomenology. An important goal will be extending
these methods to the case of multiple-field inflation.

The next immediate application will be to directly constrain parameters of
inflationary scenarios through modal bispectrum results from the \textit{Planck} satellite~\cite{Planck_NG_2018}.
The details of the work required to directly connect our coefficients to the observed data,
and the large but once-per-basis cost of this calculation, will be detailed in
a forthcoming paper~\cite{Sohn_2020}.
CMB and LSS data from forthcoming surveys will be able to use
these separable primordial bispectra to even more precisely constrain the
parameters of inflationary scenarios.

\acknowledgments
We are grateful to Wuhyun Sohn and James Fergusson for enlightening discussions and useful comments.
PC thanks the Robert Gardiner Memorial Scholarship and the Cambridge European Scholarship
for their support, through the Cambridge Trust, as well as the UK STFC.
PS acknowledges funding from STFC Consolidated Grant ST/P000673/1.

\bibliographystyle{JHEP}
\bibliography{primodal}
\end{document}